\documentclass[useAMS, usenatbib]{mn2e}
\usepackage{graphicx}
\usepackage{rotating}
\usepackage{float}
\usepackage{longtable}

\def \kmsec {\rm km~s$^{-1}$}

\def \al {\rm et al.~}
\def \13CO {$^{13}$CO}
\def \C18O {C$^{18}$O}

\def \Tmb {$T_{\rm mb}\,$}

\def \Tsys {$T_{\rm sys}$}

\def \Msun {$M_{\odot}$}

\def \arcsec {$^{\prime\prime}$~}
\def \arcmin {$^\prime$}

\def \3P1 {$^3$P$_1$ -- $^3$P$_0$}

\def \H2 {H$_{2}$}
\def \nh2 {n$_{H_2}$\,}

\begin{document}
\title{The JCMT Legacy Survey of the Gould Belt: a molecular line study of the Ophiuchus molecular cloud}
\author[Glenn J. White, Emily Drabek-Maunder, Erik Rosolowsky, Jenny Hatchell \al]
{Glenn J. White$^{1,2}$, Emily Drabek-Maunder$^{3,11}$, Erik Rosolowsky$^4$, Derek Ward-
\newauthor Thompson$^5$, C.J. Davis$^8$, Jon Gregson$^2$, Jenny Hatchell$^3$, Mireya Etxaluze$^2$,
\newauthor Sarah Stickler$^4$, Jane Buckle$^7$, Doug Johnstone$^{6,9,15}$, Rachel Friesen$^{14}$, Sarah
\newauthor Sadavoy$^9$, Kieran. V. Natt$^1$, Malcolm Currie$^9$, J. S. Richer$^7$, Kate Pattle$^5$
\newauthor Marco Spaans$^{10}$, James Di Francesco$^{6,12}$, M.R. Hogerheijde$^{13}$\\
$^1$ RALSpace, The Rutherford Appleton Laboratory, Chilton, Didcot, Oxfordshire, OX11 0QX, England\\
$^2$ Department of Physics and Astronomy, The Open University, Walton Hall, Milton Keynes, MK7 6AA, England\\
$^3$ Physics and Astronomy, University of Exeter, Stocker Road, Exeter EX4 4QL, England\\
$^4$ Department of Physics, University of Alberta, 4-181 CCIS Edmonton AB T6G 2E1, Canada\\
$^5$ Jeremiah Horrocks Institute, University of Central Lancashire, Preston, Lancashire, PR1 2HE, England\\
$^6$ National Research Council Canada, 5071 West Saanich Road, Victoria, BC, V9E 2E7, Canada\\
$^7$ Astrophysics Group, Cavendish Laboratory, J J Thomson Avenue, Cambridge CB3 0HE, England;\\
$^8$ Astrophysics Research Institute, Liverpool John Moores University, Birkenhead, Wirral, CH41 1LD, England\\
$^9$ Joint Astronomy Centre, 660 N. A`ohoku Place, University Park, Hilo, Hawaii 96720, USA\\
$^{10}$ Kapteyn Astronomical Institute, Postbus 800, 9700 AV, Groningen, Holland\\
$^{11}$ Imperial College, Blackett Laboratory, Prince Consort Road, London SW7 2AZ, England\\
$^{12}$ National Research Council Canada, 5071 West Saanich Rd, Victoria, BC, V9E 2E7, Canada\\
$^{13}$ Leiden Observatory, Leiden University, PO Box 9513, 2300 RA, The Netherlands\\
$^{14}$ Dunlap Institute for Astronomy $\&$ Astrophysics, University of Toronto, 50 St. George St., Toronto, ON, M5S 3H4, Canada\\
$^{15}$ Department of Physics and Astronomy, University of Victoria, Victoria, BC, V8P 1A1, Canada\\
}
\maketitle
\label{firstpage}
\begin{abstract}
\label{Abstract}
CO, $^{13}$CO and C$^{18}$O ${\it J}$ = 3--2 observations are presented of the Ophiuchus molecular cloud. The $^{13}$CO and C$^{18}$O emission is dominated by the Oph A clump, and the Oph B1, B2, C, E, F and J regions. The optically thin(ner) C$^{18}$O line is used as a column density tracer, from which the gravitational binding energy is estimated to be $4.5 \times 10^{39}$ J (2282~$M_\odot$~km$^2$~s$^{-2}$). The turbulent kinetic energy is $6.3~\times~10^{38}$~J (320~$M_\odot$~km$^2$~s$^{-2}$), or 7 times less than this, and therefore the Oph cloud as a whole is gravitationally bound. Thirty protostars were searched for high velocity gas, with eight showing outflows, and twenty more having evidence of high velocity gas along their lines-of-sight. The total outflow kinetic energy is $1.3 \times 10^{38}$~J (67~$M_\odot$~km$^2$~s$^{-2}$), corresponding to 21$\%$ of the cloud?s turbulent kinetic energy. Although turbulent injection by outflows is significant, but does ${\it not}$ appear to be the dominant source of turbulence in the cloud. 105 dense molecular clumplets were identified, which had radii $\sim$~0.01--0.05 pc, virial masses $\sim$~0.1--12~\Msun, luminosities $\sim$~0.001--0.1 K~km~s$^{-1}$~pc$^{-2}$, and excitation temperatures $\sim$~10--50K. These are consistent with the standard GMC based size-line width relationships, showing that the scaling laws extend down to size scales of hundredths of a parsec, and to sub solar-mass condensations. There is however no compelling evidence that the majority of clumplets are undergoing free-fall collapse, nor that they are pressure confined.
\end{abstract}


\section{Introduction}
\label{Introduction}
The Gould Belt survey is a Legacy programme that uses the James Clerk Maxwell telescope (JCMT) to map the dust and gas distributions in nearby star forming molecular clouds (Ward-Thompson \al 2007). Tilted $\sim$20 degrees from the Galactic Plane, the Gould Belt contains many of the closest low and high-mass star forming regions and OB associations. This geometry allows many of these regions to be studied at high spatial resolution, and with relatively little confusion by foreground or background Galactic emission. The goal of the present study is to map the CO, $^{13}$CO and C$^{18}$O ${\it J}$ = 3--2 distributions from the Ophiuchus Molecular cloud. Studies of other targets in the JCMT Gould Belt survey have already been presented: Orion A (Buckle \al 2012), Orion B (Buckle \al 2010), Serpens (Graves \al 2010), Taurus (Davis \al 2010), and in Perseus (Curtis \al 2010).

The paper is organised as follows: Section \ref{Section:Cloud} describes the overall Ophiuchus Cloud and previous observations, Section \ref{Section:Observations} reports the observational and instrumentation characteristics, Section \ref{Section:OphCloudResults} presents the observational data and analyses the excitation temperatures, opacities of the gas and cloud, Section \ref{Section:Outflows} determines the characteristics of the outflows and their analyses their impact on the global kinematics of the cloud, Section \ref{clump_analysis} gives the properties of the small scale structure (clumplets) across the mapped region, and Section \ref{Section:Summary} summarising the results of this study.

\section{The Ophiuchus Cloud}
\label{Section:Cloud}

Located 120$\pm$5 pc from the sun (Loinard \al 2008, Lombardi, Lada $\&$ Alves 2008), the Ophiuchus cloud is one of the closest low-intermediate mass star forming regions (Motte \al 1998, Johnstone \al 2000, Padgett \al 2008, Miville-D$\hat{e}$schenes \al 2010). It is dominated by the main Oph A clump, also known as LDN~(Lynds Dark Nebula)~1688. The line of sight extinction A$_{\rm v}$~$\sim$~50--100 mag (Wilking $\&$ Lada 1983, Khanzadyan \al 2004).

CO isotopic measurements have previously been reported by White \al (1986), Ladd (2004), Wouterloot, Brand $\&$ Henkel (2005), Minchin $\&$ White (1995) and White $\&$ Sandell (1995). A large atomic carbon, CI, and CO map of the Ophiuchus cloud has been presented by Kamegai \al (2003), and extensive $^{13}$CO measurements reported by Loren (1989a,b, 1990). Other molecular line observations have included HCO$^+$, NH$_3$, [OI], NH$_3$ by Boogert \al (2002), O$_2$ by Larsson \al (2007), Liseau $\&$ Justtanont (2009), Liseau \al ( 2012) and Maruta \al (2010), H$_2$D$^+$ by Harju \al (2008), H$_2$O by Imai \al (2007), NH$_3$ by Friesen \al (2009)), CI (Kulesa \al 2005), and in the far-IR continuum (Ceccarelli \al 1998, Liseau \al 1999a).

A bright filament of far-IR emission traces the western edge of the molecular clump, and appears associated with an edge-on photodissociation region close to HD 147889 (Liseau \al 1999a; Kulesa \al 2005, Cassassus \al 2008). Such filamentary structures are widely recognised as being associated with star forming molecular cores (Palmerin \al 2013). Mapping of the [CI] and [CII] lines show that these lines primarily trace emission from the lower density envelopes of molecular clumps illuminated from the far side of the cloud (Kamegai \al 2003; Kulesa \al 2005). This emission follows the column density distribution of the molecular gas and is modelled as having temperatures of 50--200 K, surrounded by lower temperature gas at ~$\sim$20 K (Kulesa \al 2005).

The early stages of star formation can be studied by observations of the outflows and/or jets from embedded objects, usually from CO or H$_2$ imaging (Phillips \al 1982, White \al 2000, Beckford \al 2008, Nakamura \al 2011). Previous surveys of the Ophiuchus cloud complex at millimetre, infrared and optical wavelengths have led to the identification of a number of CO molecular outflows (Loren 1989a,b, Andr$\acute{\rm e}$ \al 1990, Zhang $\&$ Wang 2009, van der Marel \al 2013, Hatchell \al 2012), as well as individual outflows associated with IRS 44 (Terebey \al 1989); VLA~1623 (Andr$\acute{\rm e}$ \al 1990, Dent \al 1995); GSS 30 (Tamura \al 1990); Oph B (Kamazaki \al 2003) and the ELIAS12/VLA~1623 region (Dent, Matthews $\&$ Walther 1995, Narayanan $\&$ Logan 2006, Bussman \al 2007), which are covered by the present observations.

Atomic jets and Herbig-Haro (HH) objects have also been discovered in the cloud (Wilking \al 1997, G$\acute{\rm{o}}$mez \al 1998, 2003, Phelps $\&$ Barsony 2004, Wu \al 2002, Smith \al 2005). Near-infrared (H$_2$ 2.12 $\micron$)~imaging observations have been used to identify shock-excited knots and jets located toward highly obscured areas (Davis $\&$ Eisl${\rm \ddot{o}}$ffel 1995; Dent \al 1995; Molinari \al 2000; Davis \al 2010; Grosso \al 2001, Habart \al 2010, 2011, Zhang \al 2013). $\it{Spitzer}$ maps of the region have been reported by Lombardi, Lada $\&$ Alves (2008), and radio continuum observations by Casassus \al (2008) and Loinard \al (2007). The distribution of pre-stellar clumps and YSOs has been presented by Ratzka \al (2005), Smith \al (2005), Young \al (2006), Stamatellos, Whitworth $\&$ Ward-Thompson (2007), J$\o$rgensen \al (2008), Enoch \al (2008), Padgett \al (2008). The turbulent nature and motions of the gas have been studied by Snow, Destree $\&$ Welty (2008), Schmeja, Kumar $\&$ Ferreira (2008) and Friesen \al (2009).

The radiation field in LDN 1688 is dominated by two embedded B-type double stars, $\rho$~Oph AB, which are located at a distance of $\mathop{110^{+12}_{-10}}$ pc, and have spectral classes B2V+B2V (Abergel \al 1996, Cassassus \al 2008, Rawlings \al 2013). These have a combined luminosity $L$~=~5800~$L_{\sun}$, with the other double system, HD 147889 at a distance of $\mathop{118^{+12}_{-11}}$ pc, and stars with spectral classes B2IV+B3IV and $L$ = 7800~$L_{\sun}$ (Rawlings \al 2013). HD 147889, is located 0.5--1 pc away from the far side of the Oph A clump. The UV-field is about 1 per cent of that of the well studied Orion Bar (White \al 2003, Liseau \al 1999a,b,  Habart \al 2003, Naylor \al 2010, Buckle \al 2012, Arab \al 2012, Bernard-Salas \al 2012). However, the interstellar dust in Ophiuchus shows unusual dust extinction properties, with components due to large dust grains (Chapman \al 2009), and very small, rapidly rotating grains (Cassasus \al 2008, Dickinson \al 2009, Castellanos \al 2011). This dust will significantly mask UV radiation resulting from star forming activity. Other early-type stars in the Upper Scorpius OB Association may also have deposited energy into the cloud (Greene $\&$ Meyer 1995, Thompson \al 2004, Luhman $\&$ Rieke 1999, Ratzka, K${\rm \ddot{o}}$hler $\&$ Leinert 2005, Zavagno \al 2010).

Although most of the condensed objects appear to be relatively old T-Tauri stars, at least six major areas undergoing active star-formation have been identified: Oph-A to Oph-F respectively, which contain many pre-stellar clumps and Young Stellar Objects (YSOs) (Loren, Wootten, $\&$ Wilking 1990, Allen \al 2002, Ridge \al 2006, Friesen \al 2009, van Kempen \al 2009). The most active star-forming part of the Ophiuchus complex, LDN~1688, contains almost half of the mass of the Ophiuchus cloud, and exhibits a relatively high gas-star conversion efficiency that has been estimated to be $\sim$22 per cent (Wilking, Lada $\&$ Young 1989). The Oph B region is the second densest clump in the Ophiuchus cloud complex, containing several 1.3 mm dust continuum sources (Andr$\acute{\rm e}$, Ward-Thompson $\&$ Barsony 1993, Motte, Andr$\acute{\rm e}$, $\&$ Neri 1998, Liseau, White $\&$ Larsson 1999b, J${\o}$rgensen \al 2008), which are likely to be very young protostars.

The above description has focussed on previous studies relevant to this paper, for more extensive details of other studies of the Ophiuchus region, the reader is referred to the compilation by Wilking, Gagne $\&$ Allen (2008).

\section{The Observations}
\label{Section:Observations}
\subsection{JCMT observations}
The HARP (Heterodyne Array Receiver Programme - see Buckle \al 2009) receiver contains an array of 16 heterodyne detectors, arranged in a 4$\times$4 footprint on the sky. HARP was used to make maps in  the CO, $^{13}$CO and C$^{18}$O ${\it J}$ = 3--2 lines, where it has a beamsize of 14$^{\prime\prime}$ at 345 GHz (corresponding to a linear size of 0.008 pc at the Ophiuchus cloud).

The HARP receptors were operated as single sideband detectors, with system noise temperatures \Tsys~$\sim$~350--500 K for the various lines. The main-beam efficiency was taken from the JCMT efficiency archive, and assumed to be $\eta_{mb}$ = 0.61 at 345 GHz; with the values of the forward coupling efficiency, $\eta_{fss}$ = 0.71, 0.72 and 0.78 at 329, 330 and 345 GHz respectively\footnote{The  C$^{18}$O, $^{13}$CO and CO ${\it J}$~=~3--2 rest frequencies were 329.3305525, 330.5879601 and 345.7959899 GHz respectively.}. In general, the data presented in this paper are in units of main beam brightness temperature $T_{\rm mb}$ (Kutner $\&$ Ulich 1981) unless otherwise stated in the text, although appropriate beam efficiency corrections were made (usually to main beam brightness temperature when discussing the compact clumps) to estimate the properties of the gas, as discussed in later sections of this paper.

The molecular line observations were made up of 3.2 hours of CO data taken in February and March 2008 and 16.6 hours of $^{13}$CO and C$^{18}$O observations taken during March, July, and August 2008. The maps were observed using the standard on-the-fly mapping mode, and referenced against an off-source reference position at RA(J2000) =  16$^h$ 38$^m$ 00.6$^s$, Dec(J2000) = -25$^{\circ}$36$^{\prime}$ 42.0$^{\prime\prime}$, which had been verified to show no line emission from examination of 60 second position-switched `stare` observation in CO. The CO data were taken with the AutoCorrelation Spectrometer and Imaging System (ACSIS) using its 250 MHz dual sub-band mode that provided 4096 channels, each with a velocity resolution $\sim$~0.05 km s$^{-1}$ per channel. The $^{13}$CO/C$^{18}$O data were taken simultaneously with each other, with each sub-band having a central rest frequency of 330.587 or 329.330 GHz respectively, providing a velocity resolution $\sim$ 0.055 km s$^{-1}$. All of the isotopologue maps were then further convolved to a resolution of 0.1 km s$^{-1}$.

To smooth out noise variations due to missing receptors or differences in receptor performance, two independent maps were made by scanning in orthogonal directions (a technique known as `basket-weaving`), which helps to mitigate scanning effects due to systematic zero level offsets between adjacent detector scans. To remove some remaining low level striping effects, the data cubes were searched for artefacts using a 3-$\sigma$ clipped image mask in each velocity channel, followed by use of a Hough transform to search for residual linear features in the data that might be indicative of such scanning artefacts aligned along the array scan direction. This destriping correction typically corresponded to cosmetic changes at the $\sim$~1--2 per cent of the peak brightness level of the map at a given point, and did not quantitatively alter the results reported here, except in a cosmetic way.

\begin{table}
\centering
{\footnotesize
\begin{tabular}{l c c c}
Region & CO & $^{13}$CO & C$^{18}$O \\
 & $T_{\rm mb}$ K & $T_{\rm mb}$ K & $T_{\rm mb}$ K \\
\hline\\
Oph B--F & 0.13 & 0.08 & 0.08 \\
Oph A & 0.19 & 0.14 & 0.11 \\
Map edges & 0.32 & 0.19 & 0.21 \\
\hline
\end{tabular}
}
\caption{The rms main beam brightness temperature noise levels for a 0.1 \kmsec~wide velocity channel in various areas of the Ophiuchus cloud for each of the isotopologues are reported in this Table.}
\label{table:noise_levels}
\end{table} 

\subsection{UKIRT observations}
To support the JCMT observations, a deep United Kingdom Infrared Telescope (UKIRT) image of the Oph region was obtained using a near-IR K-band filter, and a matching narrow-band H$_2$ image. This was used to observe the H$_2$~2.122~$\micron$~$v$ = 1--0 $S$(1) vibrationally excited line at 2.122~$\micron$~ line, which is excited in high temperature gas of several hundred up to several thousand K, and traces shocked and/or fluorescently excited material in PDRs (White \al 1987, 2000, Davis $\&$ Eisl${\rm \ddot{o}}$ffel 1995, Wilking \al 1997, G$\acute{\rm{o}}$mez \al 1998,  Dent \al  1995, Grosso \al 2001). Narrow band H$_2$ images were obtained with UKIRT during June 2010 using the near-IR wide-field camera WFCAM. This contained four Rockwell Hawaii-II (HgCdTe 2048~$\times$~2048) arrays giving an effective pixel size of 0.4$^{\prime\prime}$. Observations at four pointing positions were dithered and mosaiced together. At each of these four positions, eight exposures were obtained, each with 20 seconds integration time in the H$_2$ line, and 5 seconds in the nearby K-band continuum. The images were shifted by an integral multiple of 0.2$^{\prime\prime}$ to give a final pixel scale of 0.2$^{\prime\prime}$. The residual sky/background structure was removed by fitting a coarse surface to each image (as described by Davis \al 2010). In the final step, the broad band K-band (continuum) image was aligned with the same frame as the H$_2$ exposure, scaled and subtracted. On the final image some stars appear as negative (black) spots (due to imperfect subtraction), although regions showing diffuse emission are recovered well. Many H$_2$ emission clumps from the Ophiuchus Molecular Hydrogen Emission-Line Objects (MHOs) reported by Davis \al (2010), that were not initially visible on the UKIRT K-band continuum image, were successfully recovered using this technique, along with a number of previously uncatalogued objects. The H$_2$ image will be used later in this paper in Figures \ref{fig:OphA_br}, \ref{fig:OphB_br} and \ref{fig:OphCEF_br}.

\section{Cloud structure and physical properties}
\label{Section:OphCloudResults}

\begin{figure*}
  \centering
    \includegraphics[angle=0,width=0.92\linewidth]{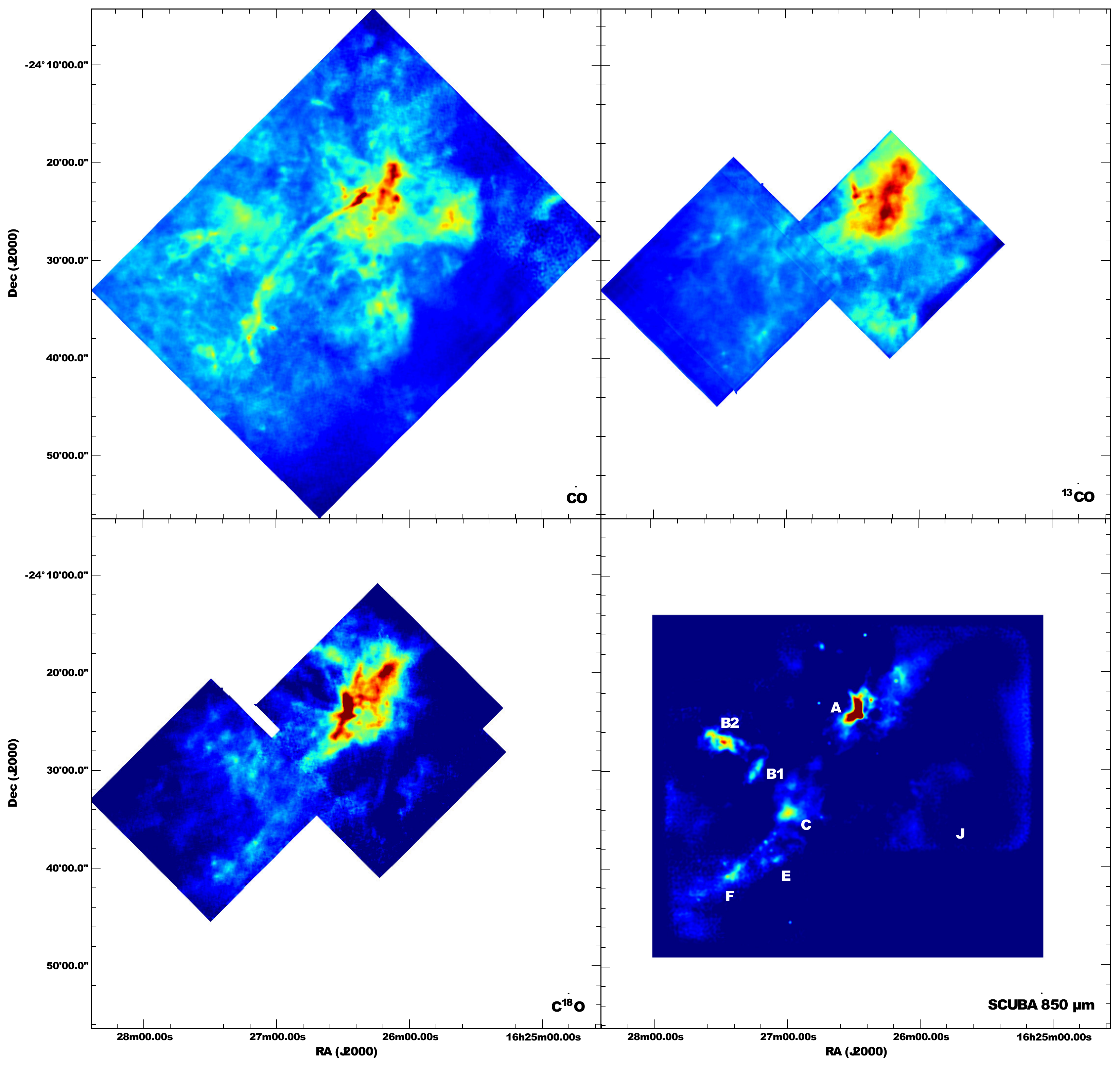}
    \includegraphics[angle=0,width=0.45\linewidth]{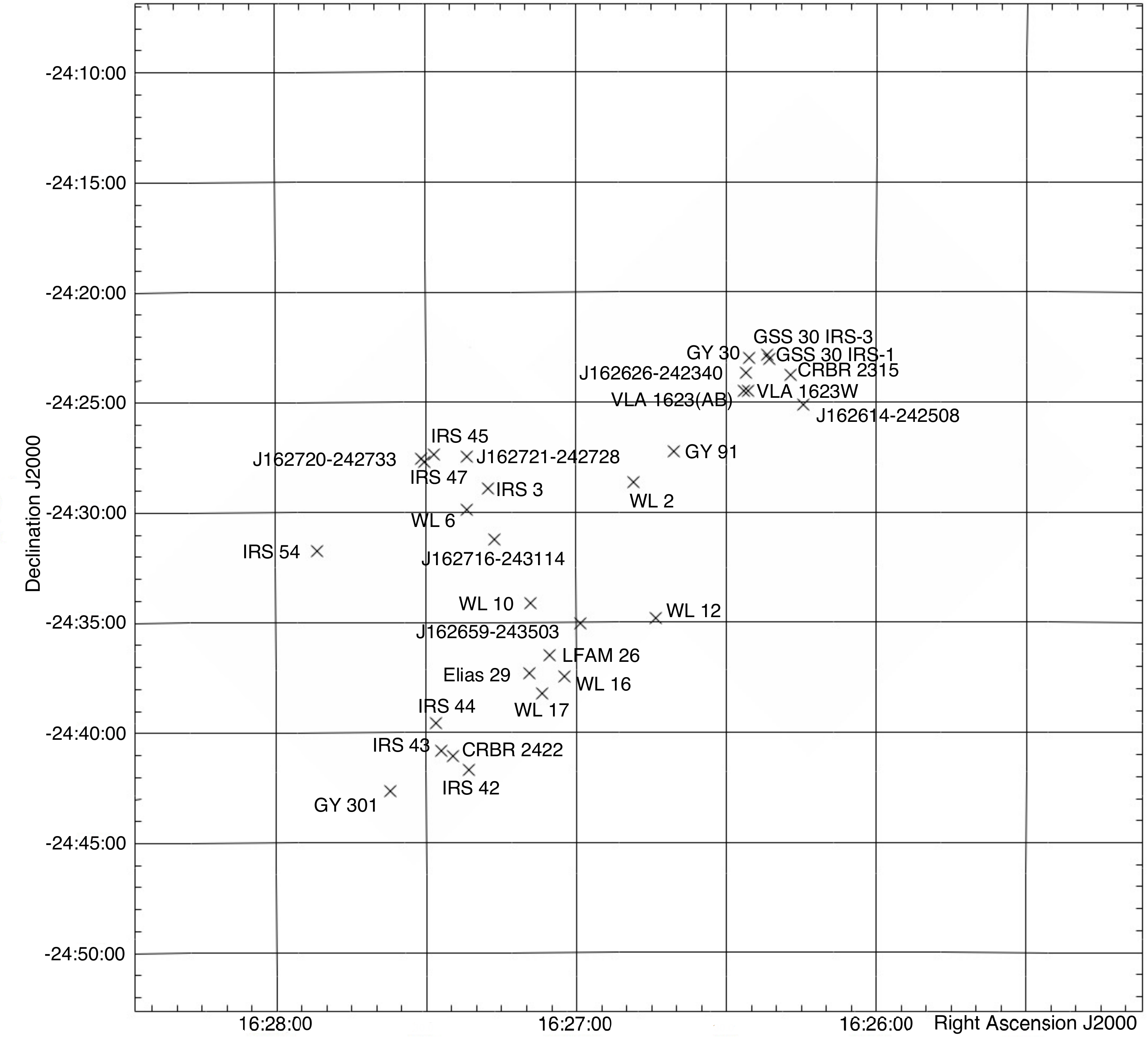}
  \caption{[Upper] CO ${\it J}$~=~3-2 observations [top left] showing the integrated emission between -5 and +12 km s$^{-1}$; [top right] $^{13}$CO emission between 0 and +7 km s$^{-1}$; [lower left] C$^{18}$O emission between 0 and +7 km s$^{-1}$ and [lower right] the SCUBA 850~$\micron$~image from Johnstone \al (2000). The maximum value in the CO ${\it J}$ = 3--2 map is T$_{mb}$ = 105 K km s$^{-1}$ close to the Oph A clump. [Lower] Orientation map showing the locations of sources referred to in the text, and in Table \ref{source_list}.}
  \label{Fig:BigCO}
\end{figure*}

For consistency throughout this paper, four kinds of cloud structure will be referred to: `cloud', meaning the overall Ophiuchus Cloud that has been observed (overall size $\sim$ several pc); `clumps', meaning individual condensations, such as Oph A, Oph B etc. with typical sizes $\sim$ 0.05--0.2 pc; sub-structures of the clumps which we call `clumplets', which refer to the smallest scale structures recovered in Section \ref{clump_analysis} from the 3-dimensional (RA, Dec, velocity) data cube, and which have typical sizes of up to $\sim$0.05 pc; and `prestellar cores', which are gravitationally bound starless cores within the cloud, which are detected by the presence of thermal continuum emission (at 1300 or 850~$\micron$). The last of these is introduced because the stability/longevity of the clumplet structures is currently uncertain (see Section \ref{clump_analysis}), and they need to be distinguished compared to the more commonly used terminology of `clumps' that feature in many other studies. The large scale integrated CO emission across the region covering Oph A and Oph B, C, E, F, and J clumps is shown in Fig \ref{Fig:BigCO}, along with the SCUBA 850~$\micron$~image from Johnstone \al (2000).

The CO map is dominated by bright emission from the Oph A region, which is itself fragmented into an hierarchy of smaller clumps. The CO map also shows the prominent curved molecular outflow that extends for $\sim$ 30\arcmin~($\sim$1 pc) away from the well studied source VLA~1623. Away from Oph A many other molecular clumps are visible, which show some correlation to the 850~$\micron$~submm continuum maps shown on the same scale. Although the~$^{13}$CO and C$^{18}$O integrated emission maps cover a smaller area than that of the CO image, they show that the isotopologues are strongly concentrated around the Oph A region, but again that many small clumps are well correlated with enhancements on the CO map, including all of those identified on the 850~$\micron$~map. Comparison with the integrated CO distribution needs to be treated with caution as many of the CO (and $^{13}$CO) lines are strongly self-absorbed by foreground emission.

\begin{figure*}
  \centering
  \includegraphics[angle=90,width=0.92\linewidth]{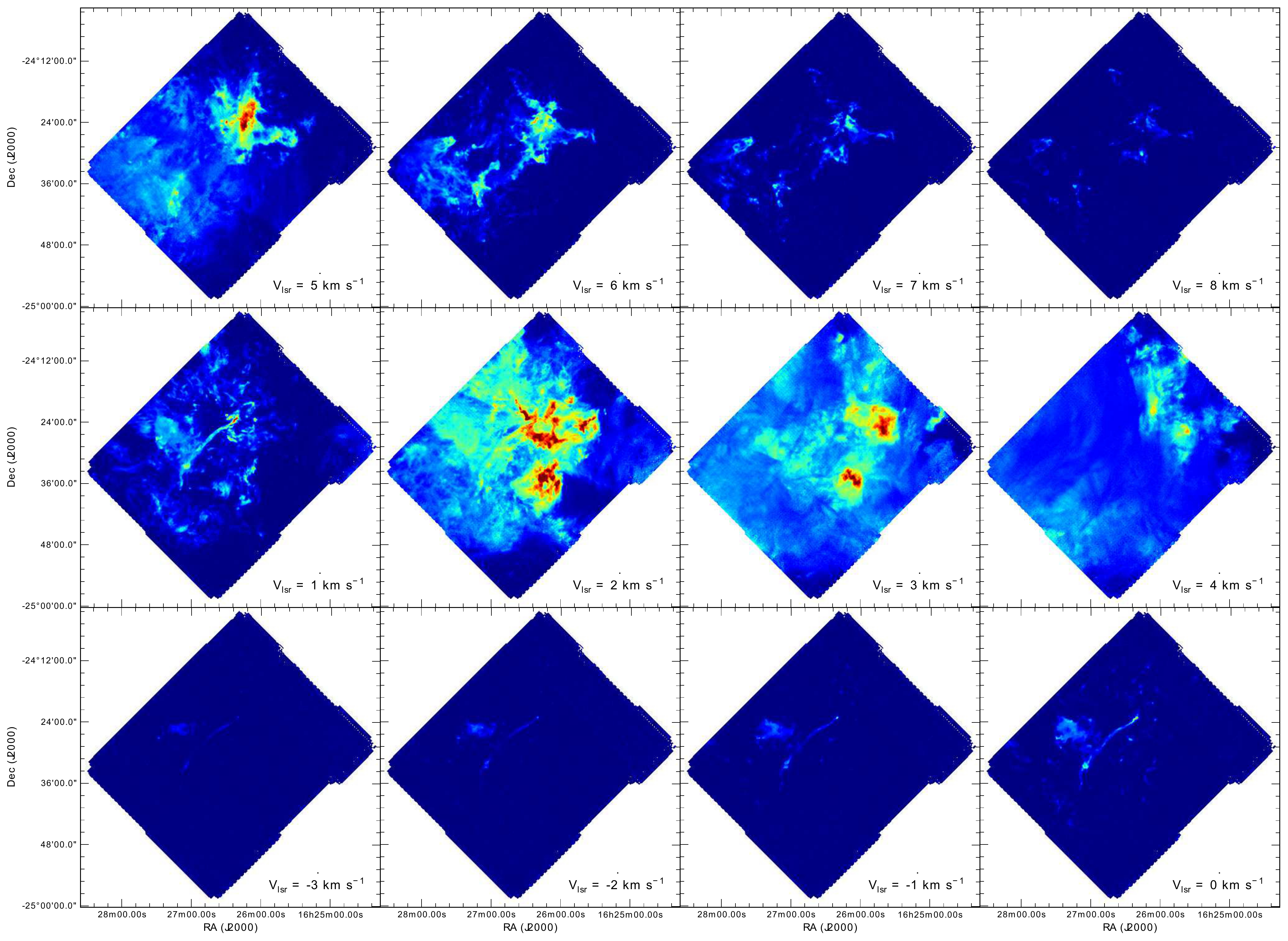}
  \caption{CO ${\it J}$ = 3--2 map in 1 km s$^{-1}$ velocity channel maps. Oph A is in the upper right hand part of the figure, and the Oph D clumps are in the lower left part of the figure. The colour scaling is shown linearly between T$_{mb}$ = 0 and 58 K.}
  \label{Fig:CompositeCO}
\end{figure*}

\begin{figure*}
  \centering
  \includegraphics[width=1.0\linewidth]{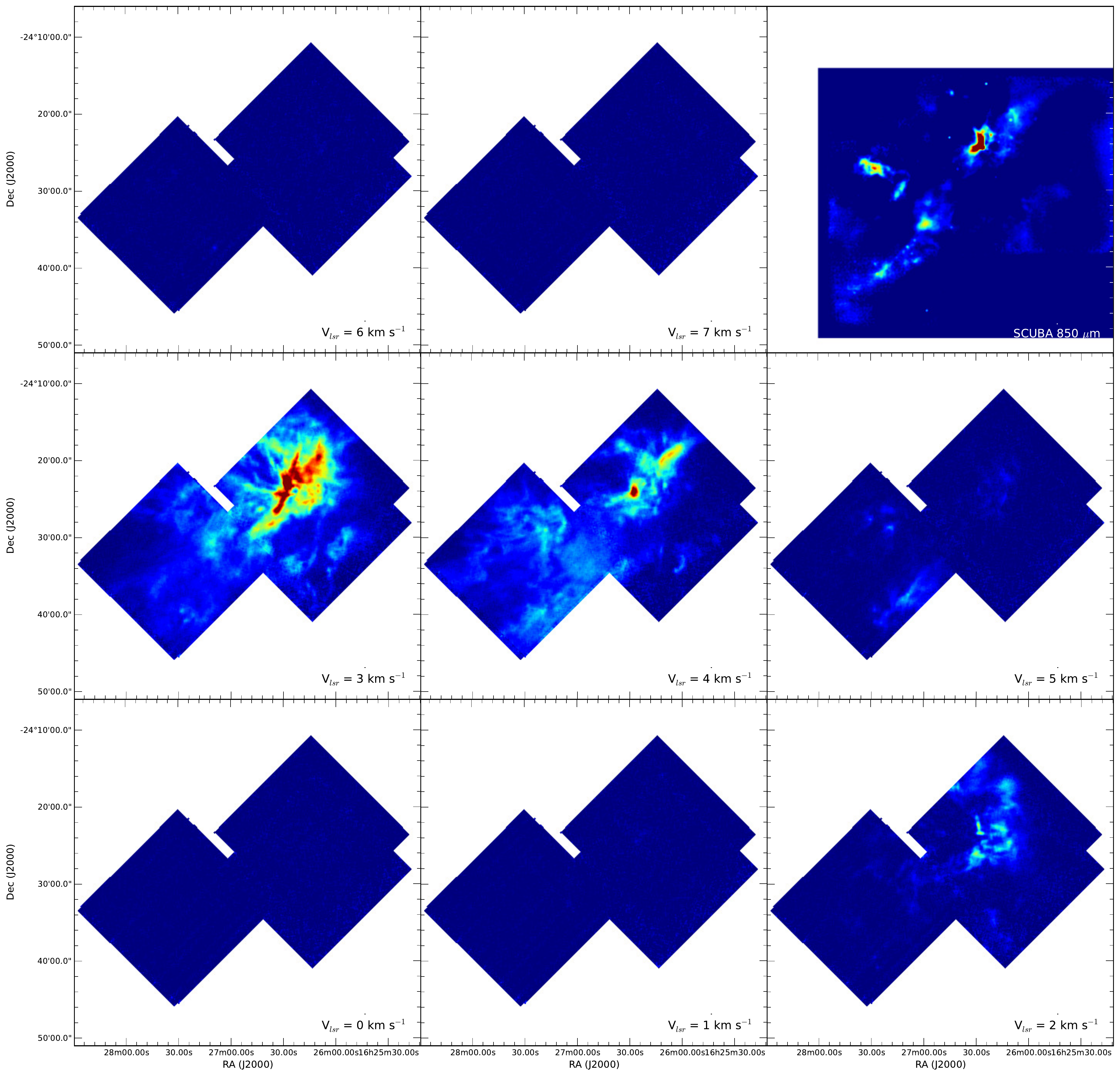}
  \caption{The C$^{18}$O ${\it J}$ = 3--2 emission between 0 (top right) and +8 (bottom left)  km s$^{-1}$. The highest value is 10.6 K km s$^{-1}$ in a 1 K km s$^{-1}$ channel close to the Oph A clump.}
  \label{Fig:C18O_velocity_channels}
\end{figure*}

The large scale CO channel maps are shown in Fig \ref{Fig:CompositeCO}. CO emission is present over a wide velocity range in excess of 12 \kmsec. From the C$^{18}$O observations (see Fig. \ref{Fig:C18O_velocity_channels}), The systemic velocity, v$_{\mathrm{lsr}}$ of the cloud, lies mostly between $\sim$~2 and 4 \kmsec, with typical C$^{18}$O half power linewidths $\sim$~1.5~\kmsec. The CO emission by contrast extends over the velocity range -3~$\rightarrow$~+8 \kmsec, with much of the gas at the higher offsets from the systemic velocity (taken from the C$^{18}$O spectra) being associated with molecular outflows. At the most blueshifted offset velocities, the maps are dominated by emission from the Oph~B2 region, and from the VLA~1623 outflow jet, the latter being seen most clearly in the +1 \kmsec~map. Bright emission from the Oph A clump region dominates at a velocity of $\sim$~+2 \kmsec, but at velocities redshifted from $\sim$ +4 \kmsec~the channel map shows little structure, particularly to the south-east of the Oph A clump, which is a consequence of the very deep self-absorption by foreground material at this velocity. At v$_{lsr}$ velocities between +5 and +8 \kmsec ~the distribution contains a lot of small-scale structure not seen in the CO maps.

\begin{figure*}
  \centering
  \includegraphics[width=1.0\linewidth]{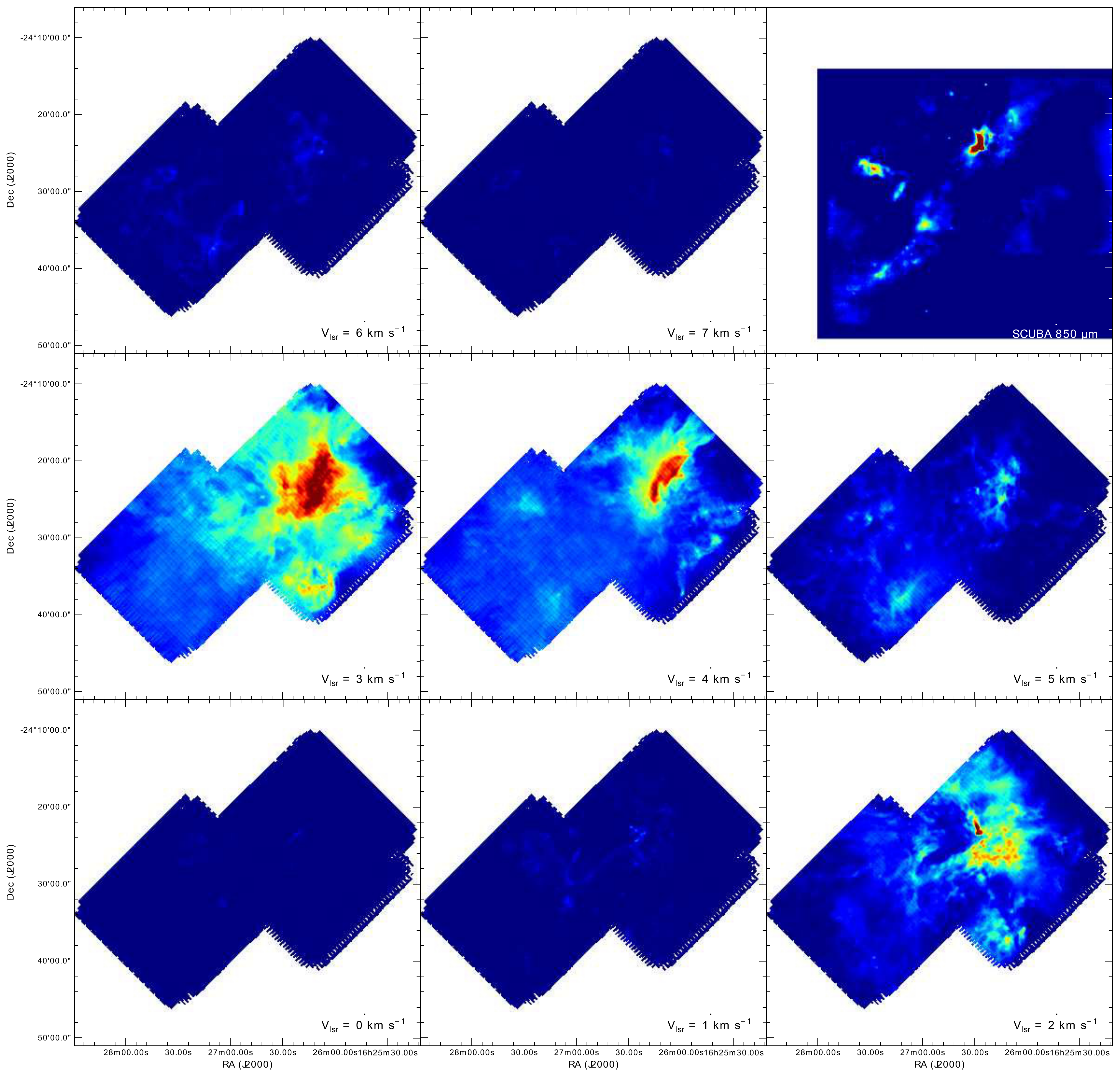}
  \caption{The $^{13}$CO ${\it J}$ = 3--2 emission between 0 (top right) and +8 (bottom left)  km s$^{-1}$. The highest value is T$_{mb}$ = 21 K km s$^{-1}$ in a 1 K km s$^{-1}$ channel close to the Oph A clump. The 850~$\micron$~map is shown again as a position reference.}
  \label{Fig:Big13CO}
\end{figure*}

 The large scale $^{13}$CO channel maps are shown in Fig \ref{Fig:Big13CO}. The $^{13}$CO emission is seen between 0 and +7 \kmsec. As with the CO channel maps, the $^{13}$CO map fails to show prominent bright small scale structure to the south-east Oph A region, even though this area contains a number of prominent submm sources. Again, this is probably a consequence of wide-spread self-absorption.

 The large scale C$^{18}$O channel maps are shown in Fig \ref{Fig:C18O_velocity_channels}. It is clear that in the velocity channels where (particularly toward the Oph A region) the CO is strongly affected by self-absorption, the C$^{18}$O channel maps show more small-scale structure, and appear to better trace the distribution of the denser material and gas clumps in the cloud. The general spatial structure from the JCMT map agrees well with that presented by Liseau \al (2010), who have mapped the C$^{18}$O ${\it J}$ = 3--2 line from an $\sim$ 10$^{\prime}$ $\times$ 5$^{\prime}$ area centred on the Oph A clump using the APEX telescope.

\begin{figure*}
  \centering
   \includegraphics[width=0.75\linewidth]{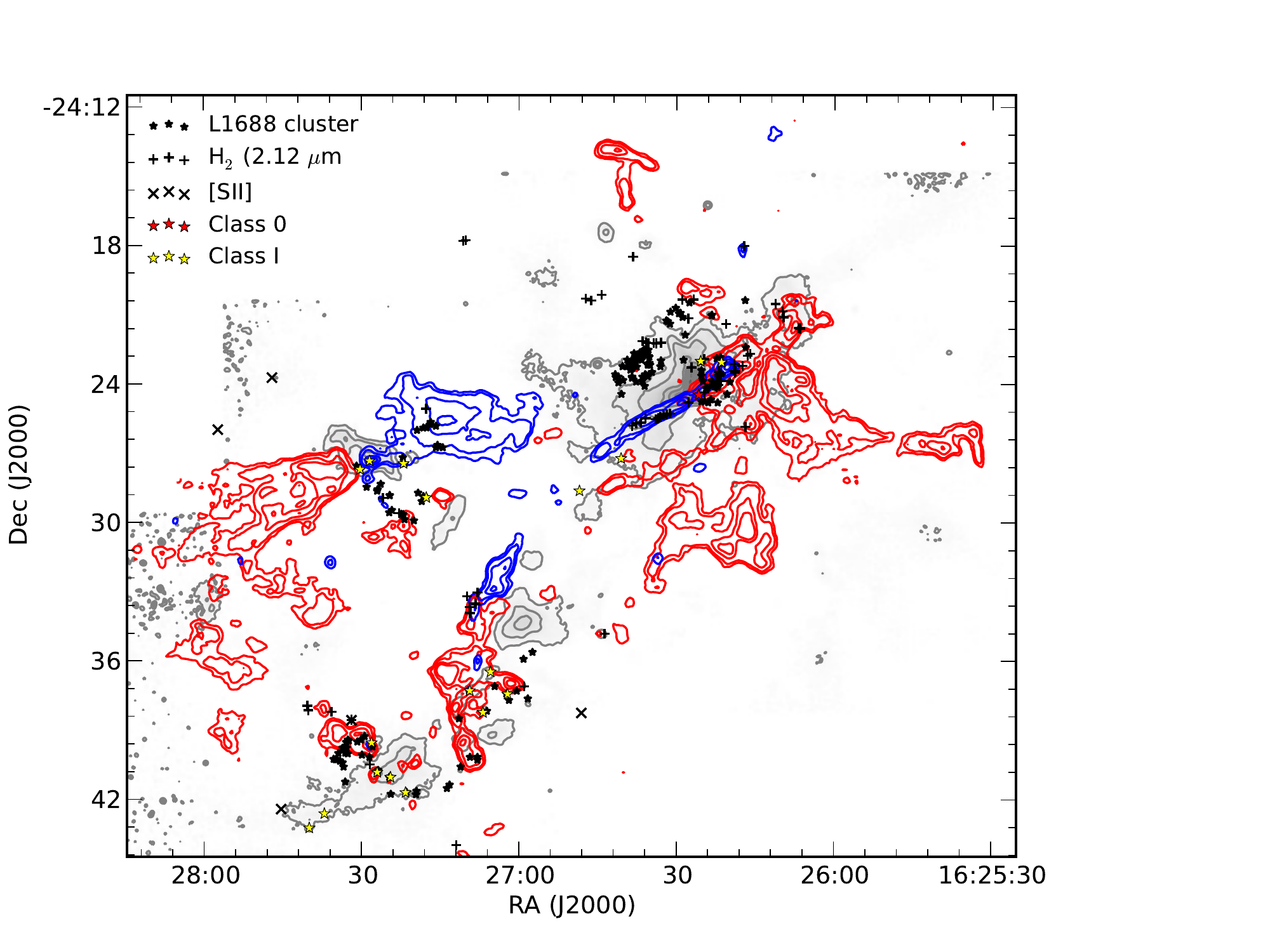}
   \includegraphics[width=0.7\linewidth]{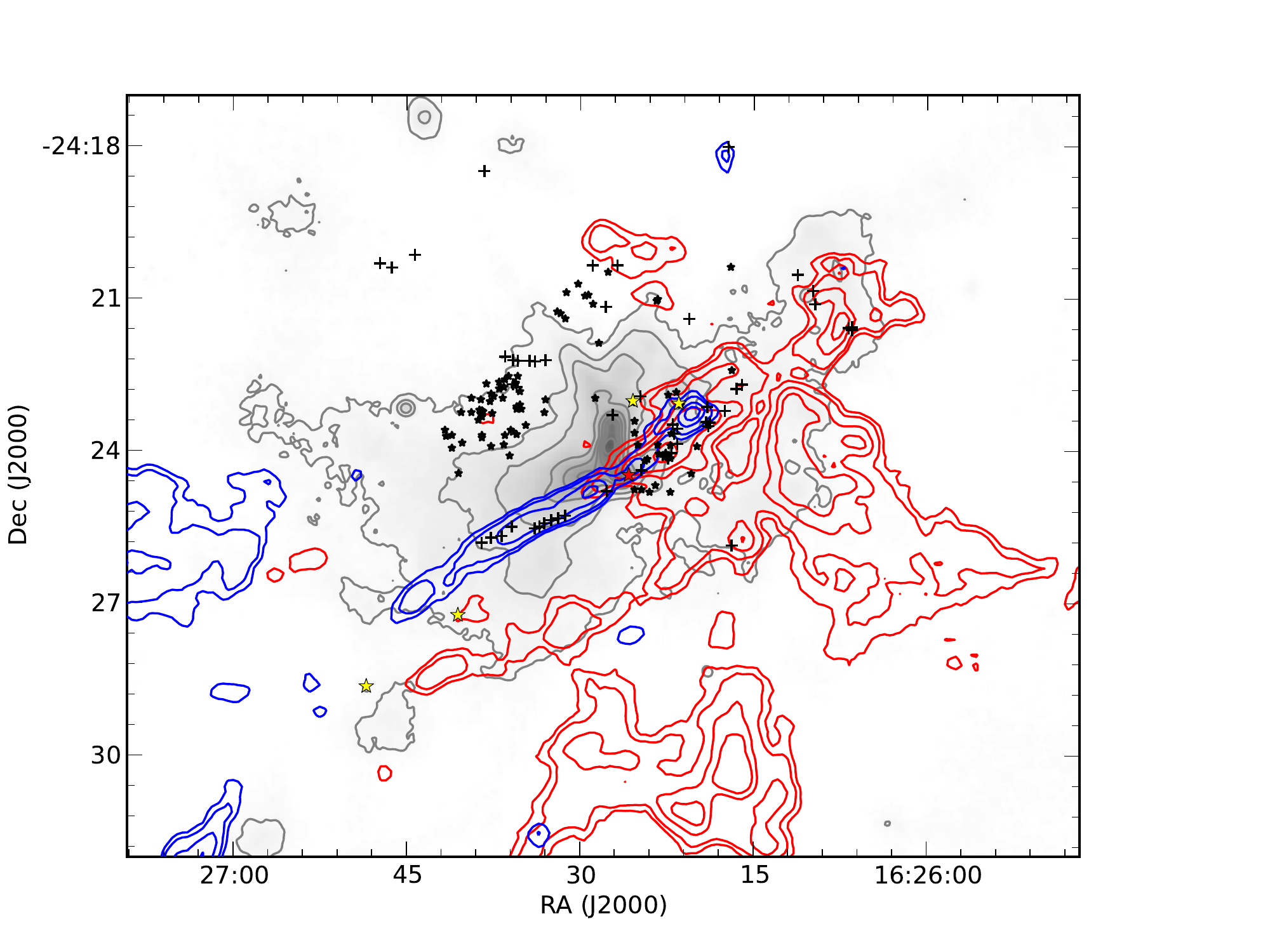}
  \caption{[Upper] Comparison between the JCMT CO data and other observations for the cloud; [Lower] Zoom-in on the Oph A clump. In both panels the blue and red outflows are appropriately colour coded, which the contour levels are arbitrarily chosen to emphasise the outflows. Blue contours are integrated from -8.6--0.4~km~s$^{-1}$ and red contours are integrated from 6.4--15.4~km~s$^{-1}$. For both maps the contour levels are at \Tmb = 3, 5, 10, 15, 30, and 45~K~km~s$^{-1}$. The annotations are from the following sources: L1688 cluster: Allen \al 2002 NICMOS, H$_2$ shocks and [SII]: Gomez \al 2003 2.12 $\mu$m data; Class 0/I protostars: Enoch \al 2009 Spitzer observations.}
  \label{Fig:12CO}
\end{figure*}

Overlays of the data with other tracers of star formation are given in Figure \ref{Fig:12CO}.

\subsection{Excitation temperatures}
\label{exc_temp}

Assuming that CO and $^{13}$CO are optically thick, excitation temperatures $T_\mathrm{ex}$ can be determined from \Tmb using Equation \ref{Eqn1}:

\begin{equation}
$$T_\mathrm{ex} = {h\nu \over k} \biggl[ \ln\biggl( {h\nu / k \over T_\mathrm{mb} + J_\nu(T_\mathrm{bg})} + 1\biggr)\biggr]^{-1}$$
{\label{Eqn1}}
\end{equation}

where $T_\mathrm{bg} = 2.73\hbox{ K}$ is the microwave background temperature and $J_\nu(T)$ is as defined in Equation \ref{Eqn2}:

\begin{equation}
J_\nu(T) = {h\nu\over k} {1\over e^{(h\nu/kT)}-1}
{\label{Eqn2}}
\end{equation}

In the Ophiuchus cloud the CO lines of the main cloud components often appear self-absorbed and cannot always reliably be used to estimate the excitation temperature, as foreground absorption by cooler gas reduces the main beam antenna temperatures. In cases where the $^{13}$CO line is optically thick, it has been be used in place of the CO line (Christie \al 2012).  The Ophiuchus region presents an exceptional case, such that even in the rare $^{13}$C$^{18}$O isotope, the lines towards some of the denser clumps also appear opaque (Liseau \al 2010).

\begin{figure*}
  \centering
  \includegraphics[width=1.0\linewidth]{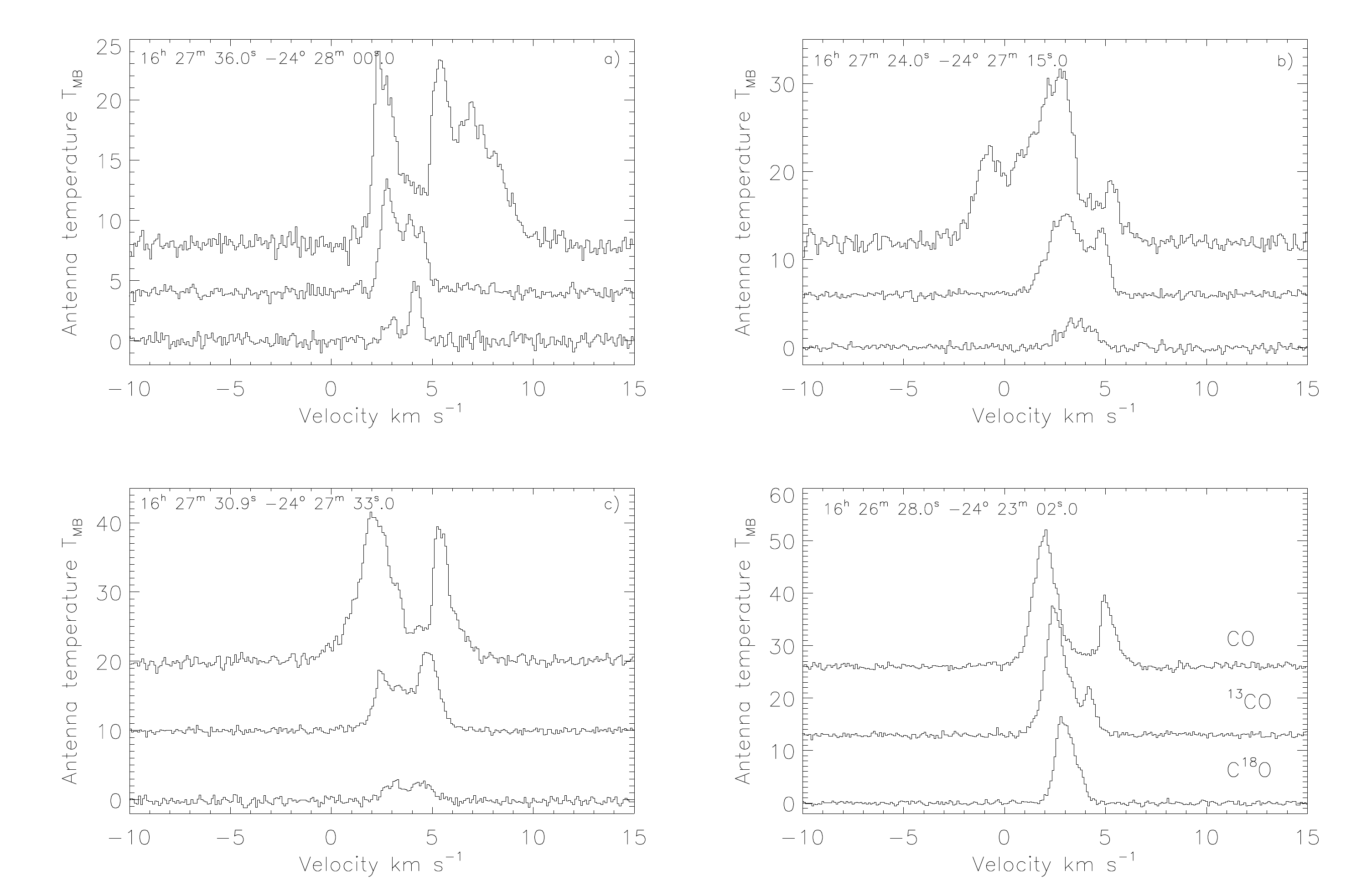}
  \caption{Typical spectra showing the data quality at two locations (each panel shows the CO, $^{13}$CO and C$^{18}$O lines ordered from top to bottom, as labelled in  panel d). Panels a)--c) show the red and blue wings of an outflow in Oph B (see Section \ref{OphB}): a) Red wing RA(J2000) = 16$^h$27$^m$36.0$^s$, Dec(J2000)= -24$^{o}$28$^{\prime}$00.0.0$^{\prime\prime}$, b) Blue wing RA(J2000) = 16$^h$27$^m$24.0$^s$, Dec(J2000)= -24$^{o}$27$^{\prime}$15.0$^{\prime\prime}$, c) Central spectrum RA(J2000) = 16$^h$27$^m$30.9$^s$, Dec(J2000)= -24$^{o}$27$^{\prime}$33.0$^{\prime\prime}$. The second position in the cloud (d) shows spectra at the peak of the C$^{18}$O emission in the Oph A clump, where self absorption is very prominent at RA(J2000) = 16$^h$26$^m$28.0$^s$, Dec(J2000)= -24$^{o}$23$^{\prime}$02.0$^{\prime\prime}$. }
  \label{COspectra}
\end{figure*}

Examples of typical spectra representative of the data quality from the Oph region are shown in Fig \ref{COspectra}. These spectra show how the spectral appearance changes amongst the various isotopologues, which is due to the widespread self-absorption that is seen throughout the cloud.

\begin{figure*}
\centering
\includegraphics[width=7.1in]{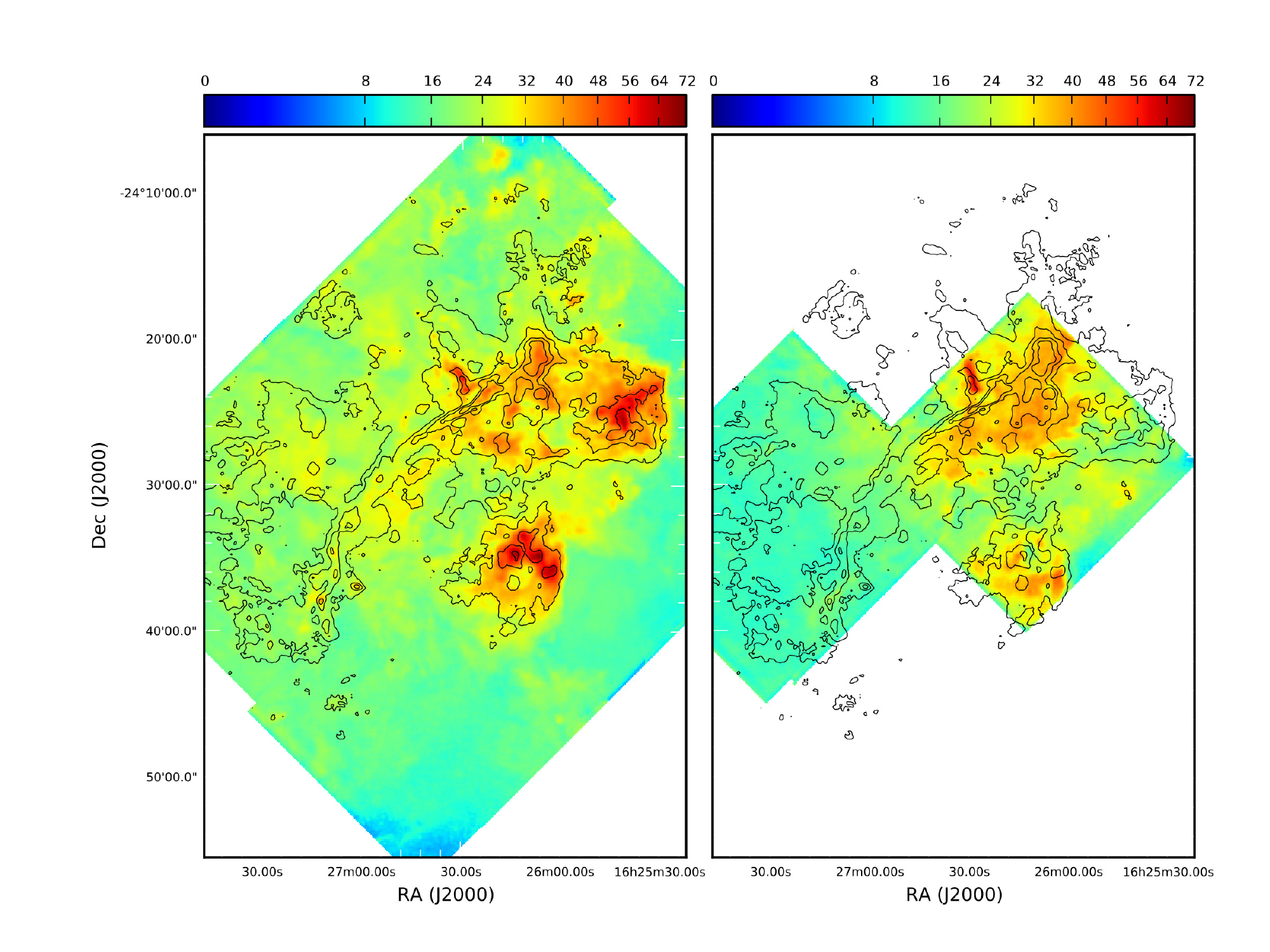}
\caption{Excitation temperature maps (K) of $T_\mathrm{ex} (\mathrm{CO})$ (left) and $T_\mathrm{ex}(^{13}\mathrm{CO})$ (right) with contours from the CO integrated main-beam intensity at levels 50, 75, 100, and 125~K~km~s$^{-1}$. }
\label{fig:excitation_12co}
\end{figure*}

The excitation temperatures for CO and $^{13}$CO estimated from the data are shown in Figure~\ref{fig:excitation_12co}.  There are strong absorption features in the CO emission ranging from $\sim$~2.0--5.5~km~s$^{-1}$ and it is possible the excitation temperature has been underestimated for this transition.  From past studies, excitation temperatures are often observed to rise in protostellar outflows (Phillips \al 1982, 1988, Rainey \al 1987, Hatchell \al 1999a, Nisini \al 2000, Giannini \al 2001).  Throughout the cloud, $T_\mathrm{ex} (\mathrm{CO})$ and $T_\mathrm{ex} (^{13}\mathrm{CO})$ are typically between $\sim$12--40~K.

A noticeable feature in the maps is the warm gas tracing the PDR close to Oph J that is illuminated by the early-type B-type stars S1 and SR3 and (mostly from) the close binary HD~147889 (B2IV, B3IV; Cassassus \al 2008), as shown in Figure \ref{Fig:CompositeCO}.  The excitation temperatures near to the PDR reach $\sim$~80~K for CO and $\sim$~50~K for $^{13}$CO.  These excitation temperatures are significantly higher than temperatures found inferred for may of the regions with molecular outflows.  For example near the VLA~1623 outflow in Oph~A, CO excitation temperatures reach 30--40~K in the blueshifted flow and higher in the redshifted flow (40--50~K), close to the PDR.  The Elias~32 and WL~6 outflows in Oph~B (see Section \ref{OphB}) have CO excitation temperatures ranging from 20--30~K, which are similar to the temperatures found in the flows from Oph~C, E, and F (including Elias~29, WL~10, IRS~44, and IRS~43 - see Sections \ref{OphCE} and \ref{OphF}).

The excitation temperatures for $^{13}$CO appear noticeably lower than that of CO even though the CO emission is affected by self absorption.  This suggests that $^{13}$CO may optically thin in some regions, and that temperature estimates based on this line should therefore be corrected to take account of this.

\subsection{Opacities in the main clumps}
\label{cloud_energetics:opacities}

In this section, the data are used to infer the optical depth of $^{13}$CO or C$^{18}$O in the cloud, and from this the mass of the cloud components. For this it is assumed that the optically thinner C$^{18}$O can be used to trace the column density and mass in the cloud.  However in a region like this suffering high extinction, it is important to confirm whether the emission is in fact optically thin.  Following Equation \ref{Eqn:CO_opacity} the optical depth is related to the intensity ratio of different CO isotopologues, $I(^{13}CO)/I(C^{18}O)$, by the relation,

\begin{equation} \frac{T_{\mathrm{^{13}CO}}}{T_{\mathrm{C^{18}O}}} = \frac{1-\exp{(-\tau{\mathrm{(^{13}CO)}})}}{1-\exp{(-\tau{\mathrm{(C^{18}O)}})}} \end{equation}

where $T$ is the brightness temperature and $\tau$ is the optical depth (of the corresponding subscripted isotopologues).  Using the method in Ladd, Fuller $\&$ Deane (1998), the ratio of $^{13}$CO and C$^{18}$O integrated intensities can be used together to provide an estimate of the optical depth of the cloud:

\begin{equation}
\frac{\int_{-\infty} ^\infty T_\mathrm{^{13}CO} (\mathrm{v}) \ \mathrm{dv}}{\int_{-\infty} ^{\infty} T_{\mathrm{C^{18}O}} (\mathrm{v}) \ \mathrm{dv}} = \frac{\int_{-\infty} ^\infty (1-\exp[-\tau_{\mathrm{^{13}CO}} (\mathrm{v})]) \ \mathrm{dv}}{\int_{-\infty} ^\infty (1-\exp[-\tau_{\mathrm{C^{18}O}} (\mathrm{v})]) \ \mathrm{dv}}
\label{eq:optical_depth}
\end{equation}

where $\tau\mathrm{(C^{18}O)} = \tau\mathrm{(^{13}CO)} / f_{18}$, with $f_{18}$ representing the abundance ratio $^{13}$CO/C$^{18}$O. 
This equation can be further simplified by setting $\tau(\mathrm{^{13}CO}) = \tau(\mathrm{^{13}CO}) \exp{[-\mathrm{v}^2 / 2\sigma^2]}$, which assumes a Gaussian velocity distribution, and predicts that the line will have a Gaussian shape.  Equation~\ref{eq:optical_depth} can then be numerically minimised to determine the optical depth, by assuming that $\sigma = 0.64$, which corresponds to the average full width half maximum (fwhm) $\mathrm{v_{fwhm}}$ of 1.5~km~s$^{-1}$ that we measure across the cloud.  The abundance ratio was assumed to be the accepted value of $X(^{13}\hbox{CO})/X(\hbox{C}^{18}\hbox{O}) = 8$ (Frerking \al 1982).  Only spectra with a peak line detection of $>$~3$\sigma_\mathrm{rms}$ were used for the $^{13}$CO and C$^{18}$O maps used to create the ratio map.

\begin{figure*}
\includegraphics[width=3.4in]{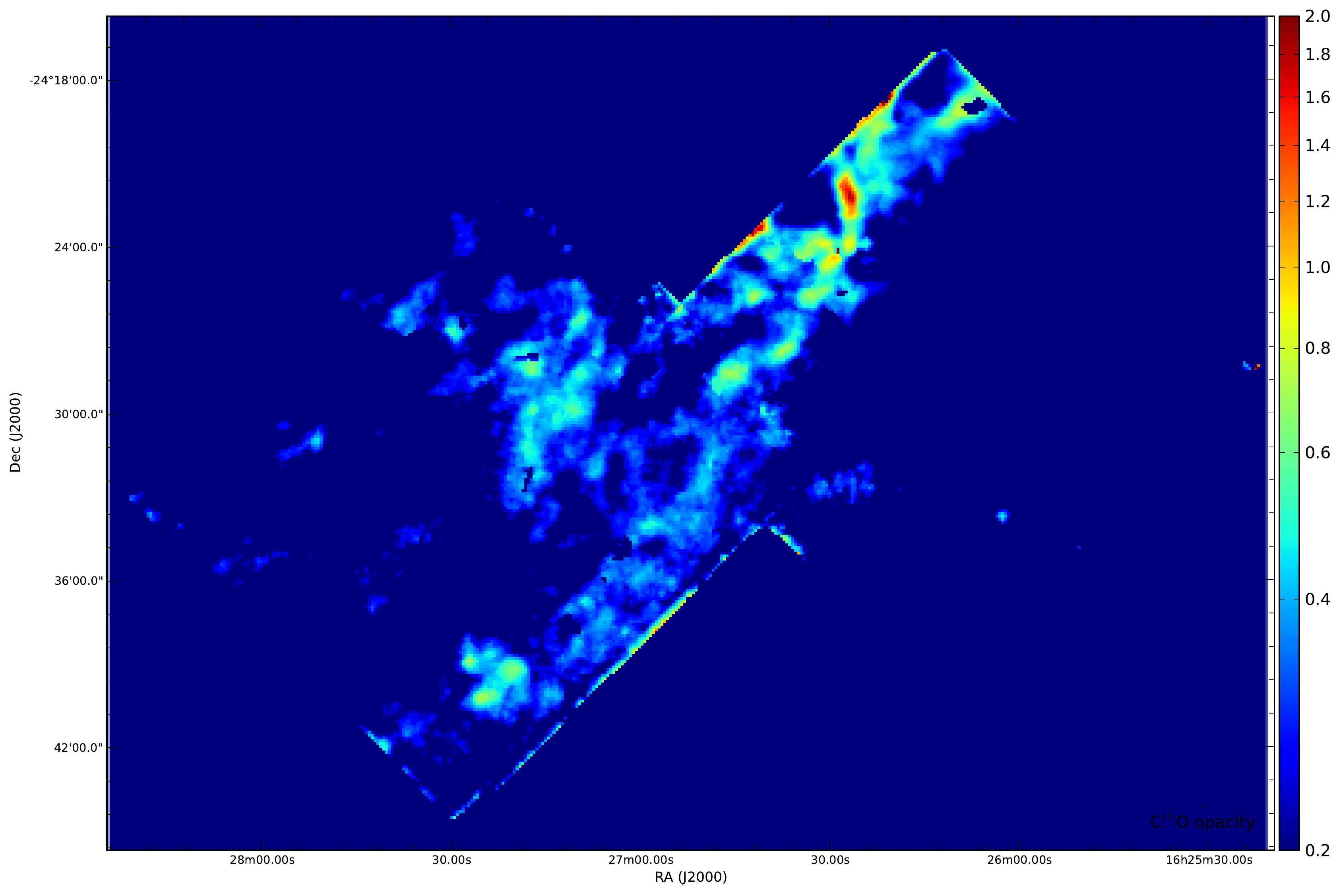}~~~~~~~~~~~~~~~~~~~~~~
\includegraphics[width=3.3in]{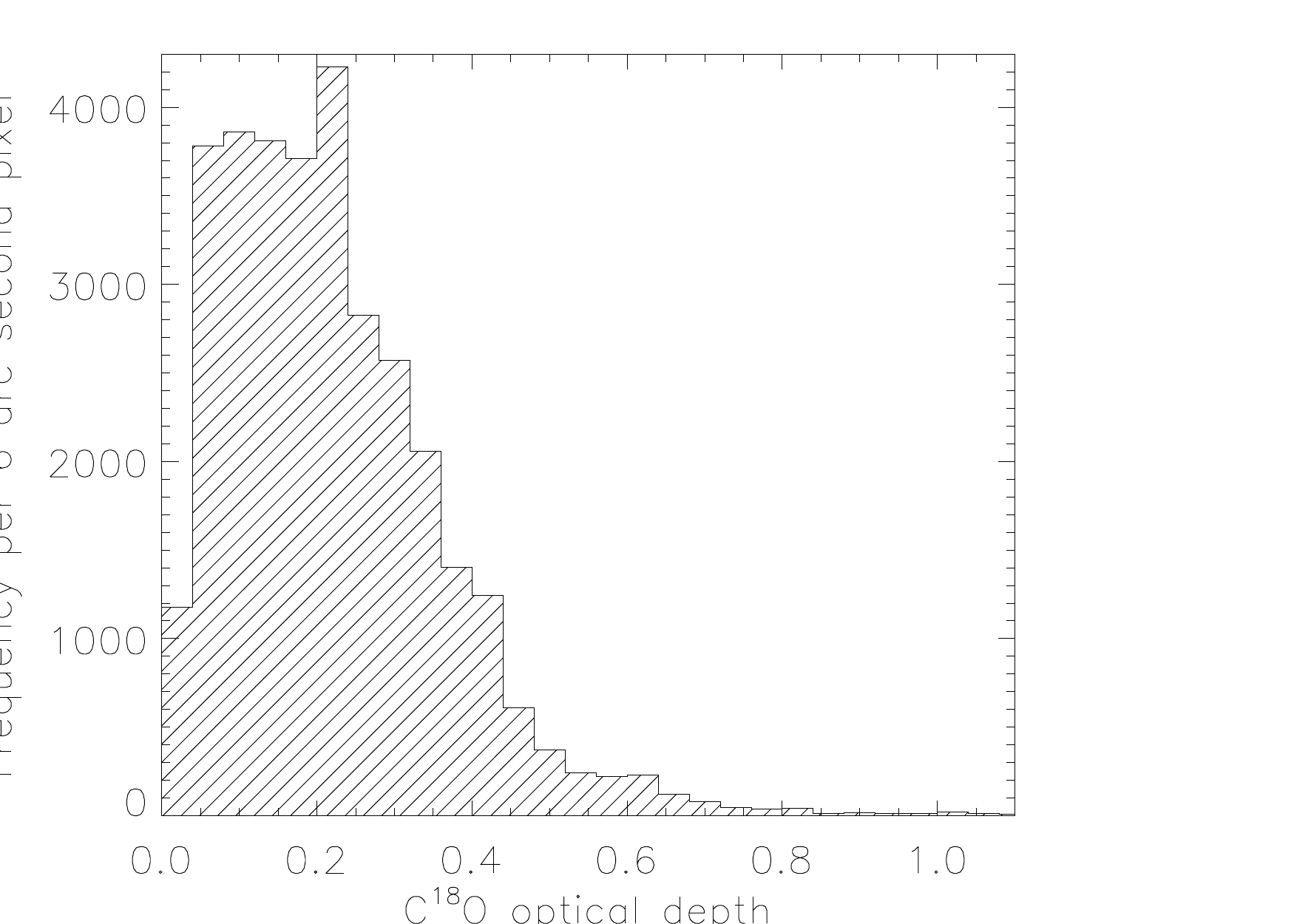}
\caption{[Left] Map of the C$^{18}$O optical depth, $\tau{(\rm {C^{18}O)}}$, following the technique described in Equations 2 and 3 of Graves \al (2010), reaching a maximum value of $\sim$~2 in the Oph A clump region. [Right] Cumulative histogram plot showing the C$^{18}$O optical depth from the map on the left. The peak C$^{18}$O optical depth is $\sim$ 0.2, with 96 per cent of the cloud area having an optical depth $<$0.5.}
\label{fig:CO_opacity}
\end{figure*}

Figure \ref{fig:CO_opacity} shows the map of $\tau{(\rm {C^{18}O)}}$, where $\tau{(\rm {^{13}CO)}}$ would be greater by the abundance factor of 8.

In Oph~A, the typical integrated intensity ratios $^{13}$CO/C$^{18}$O are observed to range from $\sim$2--7, corresponding to $\tau (\mathrm{^{13}CO})$ of 2 $\rightarrow$ 16 and $\tau(\mathrm{C^{18}O})$ of $\sim$0.3 $\rightarrow$ 2.0.  $^{13}$CO is usually found to be optically thick around the main clumps, often also showing self-absorbed profiles.  C$^{18}$O is also optically thick in some of the clumps, and toward the PDR.  In particular $^{13}$CO emission from the PDR region is self-absorbed with a $^{13}$CO/C$^{18}$O integrated intensity ratio as low as 1.2 (where $\tau (\mathrm{C^{18}O}) > 5$).

Although the optical depth of C$^{18}$O throughout most of Ophiuchus is found to be optically thin ($\tau < 1$), it is optically thick in some regions, indicating that an optical depth correction does need to be applied when calculating the mass from the integrated C$^{18}$O emission.  Including all pixels with a peak $T_\mathrm{MB}$ of at least 3$\sigma_\mathrm{rms}$ (1.36~K), the excitation temperature, $T_\mathrm{ex}$, was calculated using the $^{13}$CO peak temperature (Section~\ref{exc_temp}).  In regions where $T_\mathrm{ex}$ was degenerate, or could not be uniquely estimated, the nominal excitation temperature of $^{13}$CO was assumed to be 20~K.  The value of the abundance ratio of C$^{18}$O to H$_2$, $X_\mathrm{CO}$ was assumed to be 10$^{-7}$ (Curtis \al 2010, Graves \al 2010), which is similar to the commonly used value from Frerking \al (1982) of $1.7 \times 10^{-7}$. It should be cautioned that uncertainty remains concerning the question of whether CO abundances remain constant in molecular clouds, with recent work showing that $X_\mathrm{CO}$ may vary according to the local ISM environment, and that even within individual clouds can vary by factors of 3--4 (Lee \al 2014).

The mass of the cloud mapped in the present observations is estimated to be 515~$M_\odot$ (or without the optical depth correction 439~$M_\odot$).  Previous studies of star forming regions (Graves \al 2010, Curtis \al 2010, Buckle \al 2010) have assumed a constant excitation temperature of around 10--12~K based on $^{13}$CO average excitation temperatures.  Had the lower, constant excitation temperature been adopted, the mass would have been greater by a factor of $\sim$1.6--2.3.

\subsection{Energetics of the Ophiuchus cloud}
The virial mass of the region, M$_{\mathrm{vir}}$, were calculated under the assumption that each of the clumps in the map are spherical and virialised with a density profile of the form $\rho \propto \it{r}^{\beta}$,  where $\beta$ is the density gradient. It is further assumed that there is no magnetic support or pressure confinement, and that the size of the Ophiuchus cloud is completely enclosed by the observations. Whereas it is clear that the large-scale structure of the Ophiuchus cloud is not strictly spherical, the assumptions above serve as a first order approximation of the geometry. Subject to these uncertainties, we have used the following relationship:

\begin{equation}
M_\mathrm{vir} = {5\over \gamma} {\sigma^2 R \over G}
\label{eq:virial}
\end{equation}

where we assume that $\sigma$ is the 1D velocity dispersion $ \mathrm{\Delta v({C^{18}O})} / \sqrt{8 \ln 2}$, $R$ is the radius of the cloud in parsecs, $G$ is the gravitational constant and $\gamma$ is given by:

\begin{equation}
\gamma = -\frac{1~+~\beta/3}{1~+~2\beta/5}
\end{equation} 

which assumes that the density distributions of the clouds are spherical and follow an inverse square law. Here $\gamma$ takes the value 5/3, following that used for the Serpens (Graves et al 2010) and Orion B studies (Buckle et al 2010).

The gravitational binding energy, E$_{\mathrm{grav}}$, and the turbulent kinetic energy, E$_{\mathrm{kin}}$ can also be estimated using the relations:

\begin{equation}
E_\mathrm{grav} = -\frac{3}{5} \gamma \frac{G M^2}{R} ~~~~\rm {and} ~~~\it{E}_\mathrm{kin} = \frac{3}{2} M \sigma^2
\end{equation}

where $G$ is the gravitational constant, $M$ and $R$ are the mass and radius (assumed spherical) of the cloud respectively, and $\sigma = \mathrm{\Delta v_{C^{18}O}} / \sqrt{8 \ln 2}$ represents the 1D velocity dispersion.

Throughout the Ophiuchus cloud, the average fwhm velocity $\mathrm{\Delta v({C^{18}O}})$ was measured to be 1.5~km~s$^{-1}$ by fitting a Gaussian to the C$^{18}$O average spectrum across the cloud.  The effective radius, ${\it R}$ of the cloud was determined from $A = \pi R^2$, where $A$ is the total pixel area of the map with peak $T_\mathrm{MB}$ detections of at least 3$\sigma_\mathrm{rms}$.  This gave a radius of 14$^{\prime}$ or 0.5~pc at 120~pc distance.  Adopting these values, the virial mass is 141~$M_\odot$.

The virial mass is therefore 27~per~cent of the total estimated C$^{18}$O mass, indicating the Oph cloud as a whole is in a bound state, and most likely collapsing from an overall perspective. Our study agrees with results from past studies using $^{13}$CO~1--0 (Loren \al 1989; radius of 0.3~pc and 1.5~km~s$^{-1}$ velocity width) and C$^{18}$O~1--0 (Tachihara \al 2000; radius of 0.4~pc and 1.5~km~s$^{-1}$ velocity width) when corrected to a distance of 120~pc.
 
Loren \al (1989a,b) estimated the virial mass was 24~per~cent of the total gas mass (475~$M_\odot$ gas mass and a virial mass of 113~$M_\odot$). Tachihara \al (2000) found the virial mass to be 21~per~cent of the cloud mass (455~$M_\odot$ gas mass and a virial mass of 97~$M_\odot$ respectively).  Similarly, Nakamura \al (2011) calculated the virial mass to be 22~per~cent of the mass for the total LDN1688 cloud (R21-22, R24-26) using the $^{13}$CO data from Loren \al (1989a,b) where a distance of 125~pc and a radius of 0.8~pc was used.  Our calculated virial mass is larger than these past studies due to the higher effective radius used in Equation~\ref{eq:virial}.

\section{Molecular outflow identification and analysis}
\label{method}  
\label{Section:Outflows}

One of the objectives of this study was to investigate whether outflows in the Ophiuchus cloud are the main driving force of turbulence throughout the cloud. In this Section we first present the results from a search for molecular outflows across the mapped area, which will be followed by a discussion of the locations and interactions of these outflows with the more prominent clumps in the cloud.

Sections of the cloud close to thirty protostars selected from \emph{Spitzer} data forming part of the survey `From Molecular Clouds to Planet-Forming Disks' (c2d; Evans \al 2009) were examined to search for evidence for molecular outflows.  The c2d data was used by Padgett \al (2008) to assign a spectral index $\alpha_{2-24\micron}$ to classify the sources following earlier work by Greene \al (1994).  We further extracted those sources with a spectral index $\alpha_{2-24\micron} > -0.3$, which correspond to  Class~0 and I sources and flat spectrum objects as listed in Table~\ref{source_list}.  Sources were also required to have a bolometric temperature (Myers $\&$ Ladd 1993) ${T_{\mathrm bol}} < 750$~K to further distinguish between more evolved flat spectrum sources bordering the Class~II stage (where Class~II protostars typically have a bolometric temperature range from $650$--$2880$~K (Chen \al 1995, 1997, Andr$\acute{e}$, Ward-Thompson $\&$ Barsony 2000).  The bolometric temperature cut-off was extended to 750~K in order to include flat spectrum sources IRS~45 and 47 previously studied for driving an outflow in the Oph~B region (Kamazaki \al 2001).  The Class~II source, WL~10, was also added due to past studies (Sekimoto \al 1997) that have suggested it to be the driving source of its associated CO outflow. 

Spectra from the locations of the 30 c2d protostars were examined on the CO ${\it J}$ = 3--2 map to search these sources for line wings that could be indicative of the presence of a molecular outflow.  The method used followed the objective criterion introduced by Hatchell \al (2007), which considers whether line-wings are detected in the CO spectra above \Tmb = 1.5~K at velocity separations of $\pm 3$~km~s$^{-1}$ from v$_\mathrm{LSR}$, the ambient cloud velocity. The systemic velocities of ambient material were estimated by fitting a Gaussian to the C$^{18}$O ${\it J}$ = 3--2 spectra at the source coordinates.

C$^{18}$O velocities calculated from the Gaussian fits at the positions of the protostars can be found in Table~\ref{source_list}.  The 1.5~K temperature criterion corresponds to $T_\mathrm{mb}$~rms~values of $\sim$ 5$\sigma$ (depending on the source location in the map, after re-binning to a velocity resolution of 1~km~s$^{-1}$).  Figure~\ref{fig:12co} shows the individual CO spectra re-binned to 1~km~s$^{-1}$ with the C$^{18}$O velocities and outflow criteria marked.

The results of the outflow analysis are summarised in Table~\ref{source_list}, with the detection of either a blue and/or a redshifted line-wing noted for each source.  The line wing criterion identifies not only those protostars driving molecular outflows, but also sources which are confused by emission from nearby outflows, or which have multiple velocity components along the line of sight.  Outflow candidates identified as having a blue and/or a redshifted line-wing were further examined to determine whether the high-velocity emission detected is an outflow, and which source is the most likely driving source.  Sources with potential confusion are labelled `y?c' and the source number that could potentially cause confusion.  

To summarise the result of the outflow survey: outflows were detected toward twenty-eight out of thirty sources (eight firm associations, with twenty being more marginal - primarily due to confusion with other potential driving sources within the beam).  All of the protostars with firm outflow identifications have been previously reported in other studies (VLA~1623~AB, WL~10, Elias~29, WL~6, IRS~43, IRS~44 and IRS~54). Additionally the driver of the main Oph~B outflow (IRS~47) is discussed separately in Section~\ref{OphB}.  Several new sources identified by the c2d survey have also been examined for the presence of a molecular outflow, these are labelled `SSTc2d' in Table~\ref{source_list}.  The majority of these latter sources have some evidence for red and/or blueshifted outflow lobes, but again may be confused by other nearby protostars in this crowded region.  Non-detections are found from 2 flat spectrum sources in Oph~B and F regions.  Flat spectrum objects are less embedded than Class~0/I sources and are more likely to not have outflow detections. Each of the main regions in Ophiuchus is discussed in more detail in Sections~\ref{OphA}--\ref{OphF}.

\begin{table*}
\begin{minipage}{6.9in}
\centering
{\footnotesize
\begin{tabular}{l c c c c c c c c }
Source & Right Ascension & Declination & Name\footnote[1]{VLA: Andr$\acute{e}$ \al (1990), GY:  Greene $\&$ Young (1992), GSS:  Gradalen \al (1973), IRS:  Allen (1972), Elias:  Elias (1978), CRBR:  Comeron \al (1993), WL:  Wilking $\&$ Lada (1983), Leous \al (1991), SSTc2d:  Evans \al (2009).} & Class\footnote[2]{Classifications follow Greene \al (1994), which is used in Padgett \al (2008).} & v$_{\mathrm{LSR}}$ & Red & Blue & Verdict \\
& (J2000) & (J2000) & & & km~s$^{-1}$ & Lobe & Lobe & \\
\hline
1 & 16:26:14.63 & -24:25:07.5 & SSTc2d J162614.6-242508 &  0 & 2.8 & y & n & y?c8 \\
2 & 16:26:17.23 & -24:23:45.1 & CRBR 2315 & I & 3.1 & y & y & y?c8 \\
3 & 16:26:21.40 & -24:23:04.1 & GSS 30-IRS1, Elias 21 & I & 3.3 & y & y & y?c8 \\
4 & 16:26:21.70 & -24:22:51.4 & GSS 30-IRS3, LFAM 1 & I & 3.3 & y & y & y?c8 \\
5 & 16:26:25.46 & -24:23:01.3 & GY 30& I & 3.3 & y & y & y?c8 \\
6 & 16:26:25.62 & -24:24:28.9 & VLA 1623W \footnote[3]{Enoch \al (2009) denote this source as VLA~1623.} & I 
& 3.1 & y & y & y?c8 \\ 
7 & 16:26:25.99 & -24:23:40.5 & SSTc2d J162626.0-242340 & flat & 3.2 & y & y & y?c8 \\
8 & 16:26:26.42 & -24:24:30.0 & VLA 1623(AB) & 0 & 3.4 & y & y & y \\
9 & 16:26:40.46 & -24:27:14.3 & GY 91 & flat & 3.0 & y & n & y?PDR \\
10 & 16:26:44.19 & -24:34:48.4 & WL 12 & I & 4.0 & y & n & y? \\ 
11 & 16:26:48.47 & -24:28:38.7 & WL 2, GY 128 & flat & 3.2 & n & y & y?c4 \\
12 & 16:26:59.10 & -24:35:03.3 & SSTc2d J162659.1-243503& I & 3.8 & y & n & y? \\
13 & 16:27:02.32 & -24:37:27.2 & WL 16, GY 182 & I & 4.1 & n & y & y? \\
14 & 16:27:05.24 & -24:36:29.6 & LFAM 26, GY 197 & I & 4.3 & y & y & y?c17 \\
15 & 16:27:06.75 & -24:38:14.8 & WL 17, GY 205 & I & 4.3 & n & y & y?c17 \\
16 & 16:27:09.09 & -24:34:08.3 & WL 10 & II & 3.4 & y & y & y \\
17 & 16:27:09.40 & -24:37:18.6 & Elias 29, GY 214 & I & 4.3 & y & y & y \\
18 & 16:27:16.39 & -24:31:14.5 & SSTc2d J162716.4-243114 & flat & 3.5 & n & n & n \\
19 & 16:27:17.58 & -24:28:56.2 & IRS 3, GY 2447 & I & 4.1 & n & y & y?c27 \\ 
20 & 16:27:21.45 & -24:41:43.0 & IRS 42, GY 252 & flat & 3.8 & y & n & y? \\
21 & 16:27:21.79 & -24:29:53.1 & WL 6 & I & 3.8 & y & y & y \\
22 & 16:27:21.82 & -24:27:27.6 & SSTc2d J162721.8-242728 & flat & 3.7 & n & y & y?c27 \\
23 & 16:27:24.58 & -24:41:03.1 & CRBR 2422 & I & 4.0 & y & n & y?24 \\
24 & 16:27:26.91 & -24:40:50.7 & IRS 43, GY 265 & I & 3.7 & y & y & y \\
25 & 16:27:27.99 & -24:39:33.4 & IRS 44, GY 269 & I & 3.8 & n & y & y \\
26 & 16:27:28.44 & -24:27:21.0 & IRS 45, Elias 32 & flat\footnote[4]{IRS 45 was listed as a Class II protostar by Wilking \al (2001). However its spectral index (Greene \al 1994) suggests that it may instead be a flat spectrum source, although its bolometric luminosity lies close to the boundary between Class I and II protostars.} & 3.5 & n & y & y?c27 \\
27 & 16:27:30.17 & -24:27:43.7 & IRS 47, IRS 47 & flat & 4.0 & n & y & y \\
28 & 16:27:30.91 & -24:27:33.2 & SSTc2d J162730.9-242733 & I & 3.8 & y & y & y?c27 \\
29 & 16:27:37.23 & -24:42:37.9 & GY 301 & flat & 3.7 & n & n & n \\
30 & 16:27:51.79 & -24:31:45.4 & IRS 54, GY 378, YLW 52 & flat & 4.0 & y & y & y \\
\hline
\end{tabular}
}
\end{minipage}
\caption{Outflow status for sources in Ophiuchus based on the criteria of outflow identification, discussed in Section~\ref{method}.}
\label{source_list}
\end{table*}

\subsection{Oph~A}
\label{OphA}

The Ophiuchus cloud has been estimated to have a mass $\sim$~550 \Msun, with the denser material distributed across 14 smaller clumps with diameters from 0.05--0.3 pc (Motte, Andr$\acute{\rm e}$ $\&$ Neri 1998) with masses from $\sim$~1--44 \Msun, which together have a total mass $\sim$ 200 \Msun. These authors identified 59 submm sources associated with starless cores, with diameters between $\sim$ 1000-17,000 AU, and masses in the range 0.1--1.5 \Msun.

\begin{figure}
\centering
\includegraphics[width=3.3in]{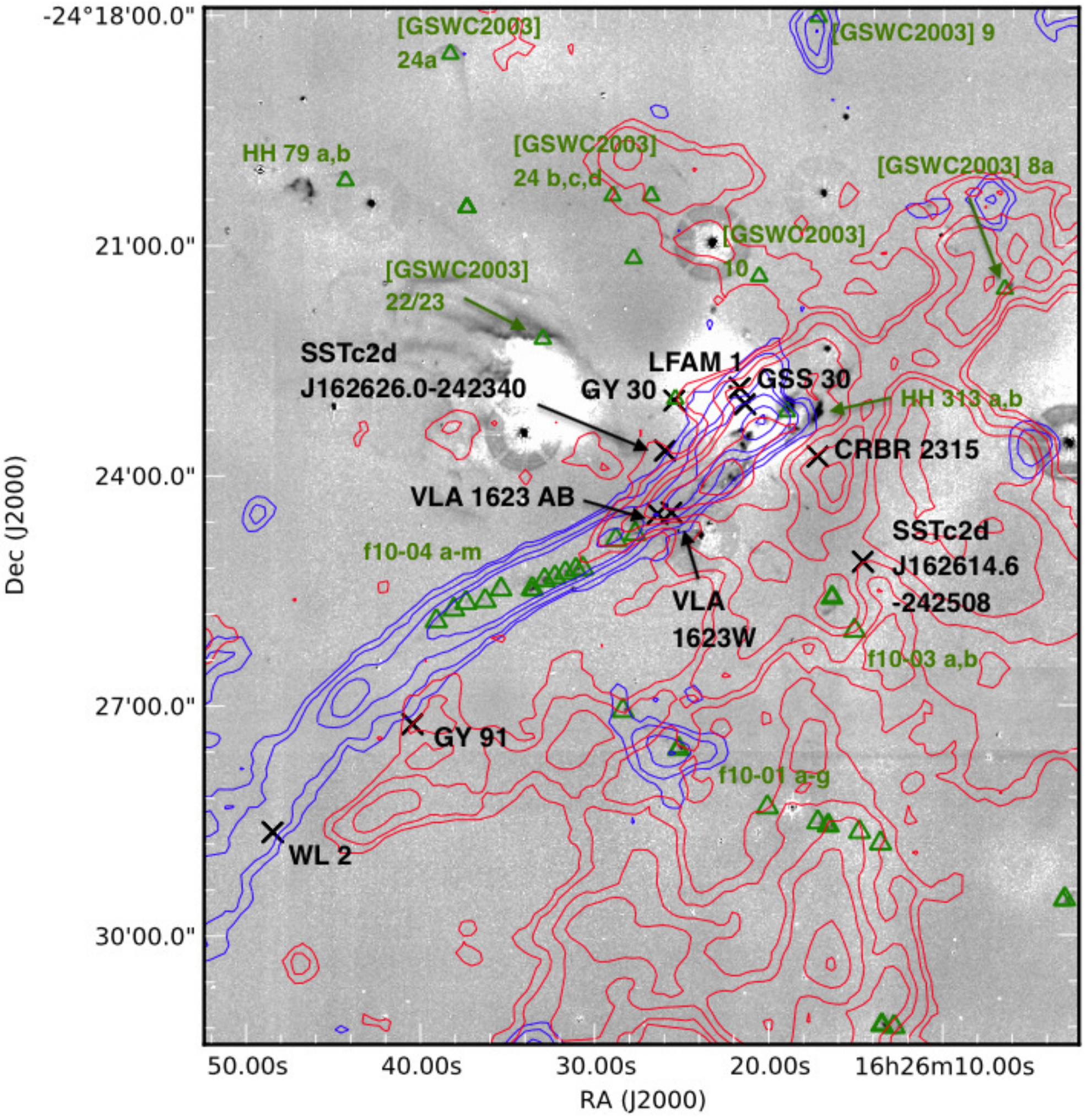}
\caption{CO~${\it J}$ = 3--2 outflows in Oph~A, superimposed onto the UKIRT continuum-subtracted H$_2$ image. Blue contours are integrated from -8.6--0.4~km~s$^{-1}$ and red contours are integrated from 6.4--15.4~km~s$^{-1}$.}
\label{fig:OphA_br}
\end{figure}

Oph~A contains two Class~0, five Class~I, and three flat spectrum protostars, dominated by VLA~1623~AB, and with potential outflows from SSTc2d~J162614.6-242508, CRBR 2315, GSS~30, LFAM~1, GY~30, VLA~1623~W, SSTc2dJ1626.0-242340 and GY~91 with at least one high velocity outflow lobe.  Figure~\ref{fig:OphA_br} shows the H$_2$~2.122~$\micron$~$v$ =  1--0 $S$(1) ro-vibrational line image of the Oph~A region with contours of the red and blueshifted CO outflows overlaid.

The highly collimated outflow from VLA~1623 can be clearly seen in Figure~\ref{fig:OphA_br}.  The prominent blueshifted lobe tracing the outflow extends roughly 0.7~pc across Oph~A.  VLA~1623 is composed of three sources (Murillo $\&$ Lai 2013, Ward-Thompson \al 2011): VLA~1623~A (Class 0), VLA~1623~B (suggested to be either between prestellar and Class~0 stages, or a shocked knot in the jet) and VLA~1623~W (Class~I), and is reminiscent of the multiple component protostellar system in the Lynds dark cloud LDN 723 (Avery, Hayashi $\&$ White 1990).  We refer to VLA~1623~A and B as one source because it is unresolved in the c2d survey.  Previous work (Murillo $\&$ Lai 2013, Dent $\&$ Clarke 2005) suggests that two distinct outflows make up the highly collimated Oph~A outflow, potentially driven by VLA~1623~A and W.  Due to confusion in the region, VLA~1623~AB is listed as the driver of this outflow, although in reality the situation appears to be much more complex, with recent 690 GHz submillimetre observations with ALMA revealing that the source IRAS 16293 B is also driving a south-east compact outflow (Loinard et al. 2013). The less prominent redshifted lobe from VLA~1623 to the north-west causes confusion in this region for Class~I sources GY~30, GSS~30, LFAM~1,  CRBR~2315, VLA~1623W and Class~0 source SSTc2d~J162626.0-242508.  Each of these sources have both blue and red line-wing detections (SSTc2d~J162626.0-242508 has a detection only in the red line-wing) using the CO outflow criterion.  There is a possibility that H$_2$ knots in this region could be driven by outflows from these sources.   GSS~30 and LFAM~1 have previously been noted as potential drivers for HH~313~b (Caratti o Garatti \al 2006), [GSWC2003]~22a and 23c (G$\acute{\rm{o}}$mez \al 2003), [GSWC2003]~9 and 10 (G$\acute{\rm{o}}$mez \al 2003),  [GSWC2003]~24b, 24c, and 24d (G$\acute{\rm{o}}$mez \al 2003).  

Similarly, high velocity gas was also seen towards GY~91 and WL~2, although whether these particular objects are the driving force of an outflow is not clear.  These sources are located in the south of Oph~A near the collimated blueshifted lobe of the VLA~1623 outflow.  Each of these sources only has one high velocity component identified in the CO molecular outflow analysis, and neither source appears to show a bipolar outflow morphology in Figure~\ref{fig:OphA_br}.  The blueshifted line-wing for WL~2 is most likely the result of the VLA~1623 outflow and the redshifted line-wing for GY~91 may be caused by other velocity components (i.e. by stellar winds from the nearby B2V stars).  In addition to these sources, some high velocity blueshifted emission can be seen in this region directly corresponding to H$_2$~knots f10-01~a--g Khanzadayan \al (2004), expected to be driven by Class~II source YLW~31 (Bontemps \al 2001).  

\subsection{Oph~B}
\label{OphB}

\begin{figure*}
  \centering
    \includegraphics[width=1.0\linewidth]{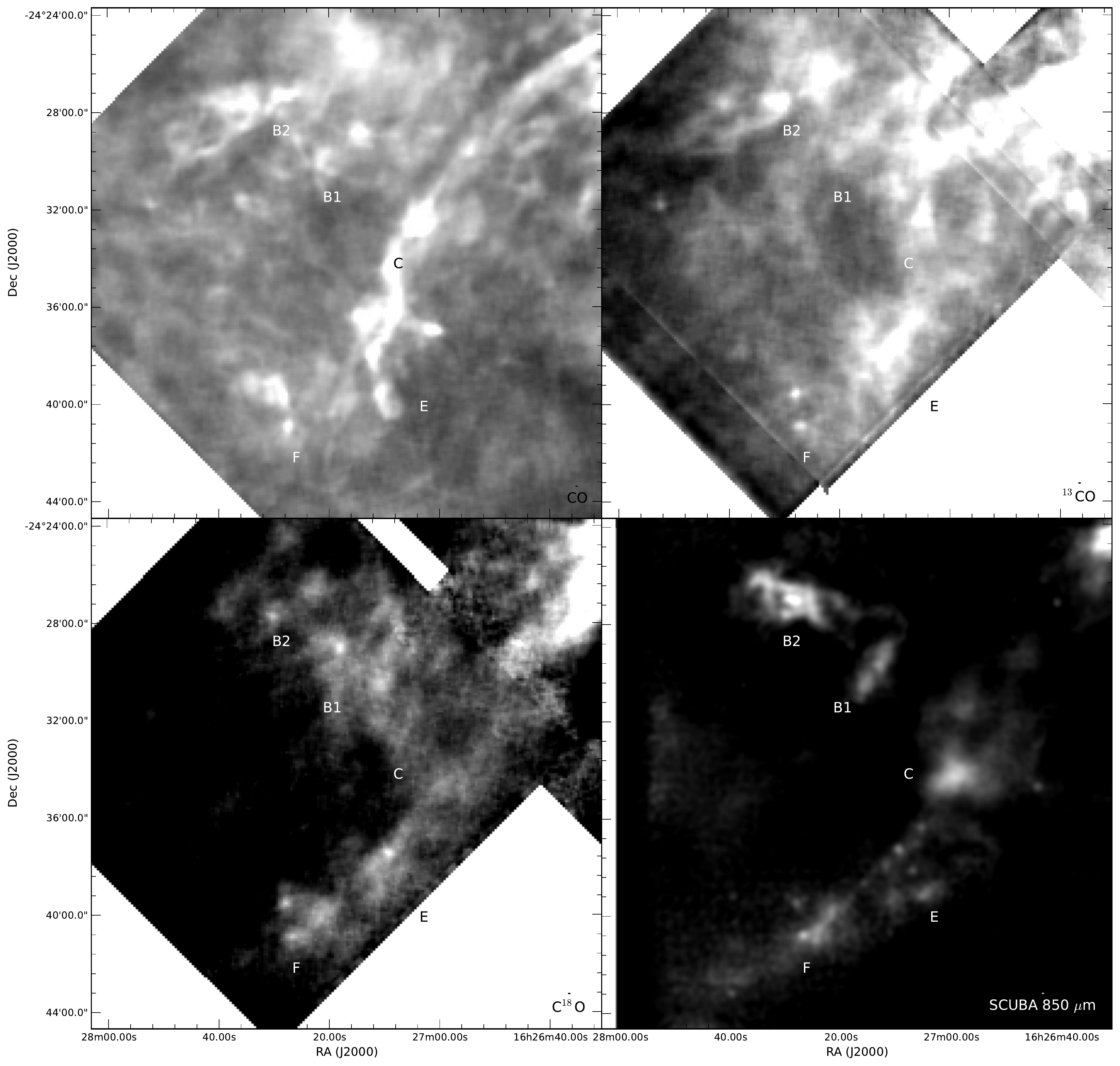}
  \caption{Greyscale plot of the Oph B, C, E and F clumps, where the three isotopologues are shown side by side, and on the same scale, to compare and contrast the appearance in the four isotopomeric lines. They figures are displayed in black and white to emphasise the changing structures between the isotopologue maps, and cosmetically saturated to emphasise the small scale structure in the data. To allow comparison the labels on the plots hold the same positions from line to line to set points of reference to guide the eye. Because of the non-linear transfer function used to enhance the features, some residual striping and joins between overlapping data blocks remain visible in the $^{13}$CO data at the few percent level, although this is at a very low level.}
  \label{Fig:c18o_cores}
\end{figure*}

The Oph B region contains two distinct active centres, Oph B1 and Oph B2. These are separated by $\sim$5 arc minutes, and clearly shown in each of the isotopomeric lines shown in Figure \ref{Fig:c18o_cores}, which shows a grey-scale plot of the region to the south-east that contains the Oph~B clumps, along with clumps C, E and F. This shows the importance of mapping complex regions like this in a number of isotopologues as well as in the submm continuum, because of the effects of opacity on changing the apparent structures. For example, many of the small submm clumps seen in the 850~$\micron$~image between Oph C, E, and F are also recovered in $^{13}$CO and C$^{18}$O, but may not be identified from the CO observations alone.

Oph B2 is the second densest region of LDN1688, and contains a number of 1.3 mm continuum sources that do not have infrared counterparts, and which are protostellar candidates (Motte, Andr${\rm \acute{e}}$ $\&$ Neri 1998, Friesen \al 2009). The Oph B2 clump appears to be rather colder than Oph A, with temperatures ranging from $\sim$ 7--23 K at the centre of the clump (Stamatellos \al 2007), and with an outflow which has previously been believed to be powered by, or associated with the Class II object IRS-45 (also known as VSSG 18 or Elias 32), or IRS-47 (van der Marel \al 2013).

The Oph-B region contains at least three Class~I and five flat spectrum protostars.  Three sources (WL~6, IRS~54, and IRS~47) were confirmed to drive outflows with four additional sources being marginal molecular outflow candidates.
 
\begin{figure}
\centering
\includegraphics[width=3.35in]{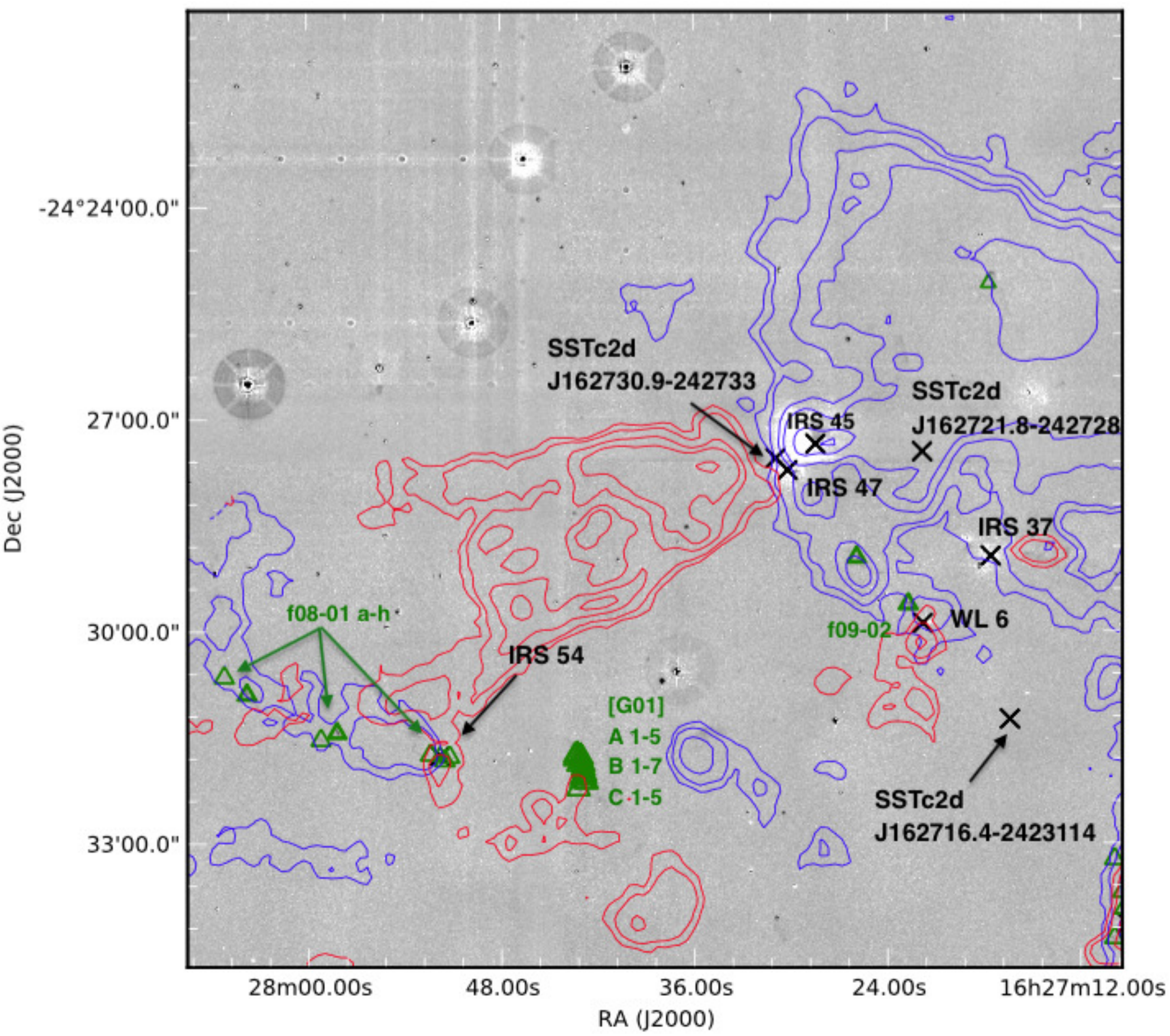}
\caption{CO~${\it J}$ = 3-2 outflows in Oph~B. Blue contours trace the blueshifted CO intensity $\int T_{\mathrm{MB}} \ {\mathrm{dv}}$ (integrated from $-6.0$ to $1.0$~km~s$^{-1}$) and red contours trace the redshifted CO intensity (integrated between 7.0 and 14.0~km~s$^{-1}$); contours are at 3, 5, 10, 15, 30, and 45~K~km~s$^{-1}$.}
\label{fig:OphB_br}
\end{figure}

Figure~\ref{fig:OphB_br} shows the H$_2$ map with the outflows for this region overlaid as contours.  The blue contours trace the blueshifted integrated CO intensity $\int T_{\mathrm{MB}} \ {\mathrm{dv}}$ (from $-6.0$ $\rightarrow$ $1.0$~km~s$^{-1}$) and the red contours trace the redshifted CO) integrated intensity (from 7.0 $\rightarrow$ 14.0~km~s$^{-1}$). For both the contours are at 3, 5, 10, 15, 30, and 45~K~km~s$^{-1}$.

\begin{figure*}
\centering
\includegraphics[width=3in]{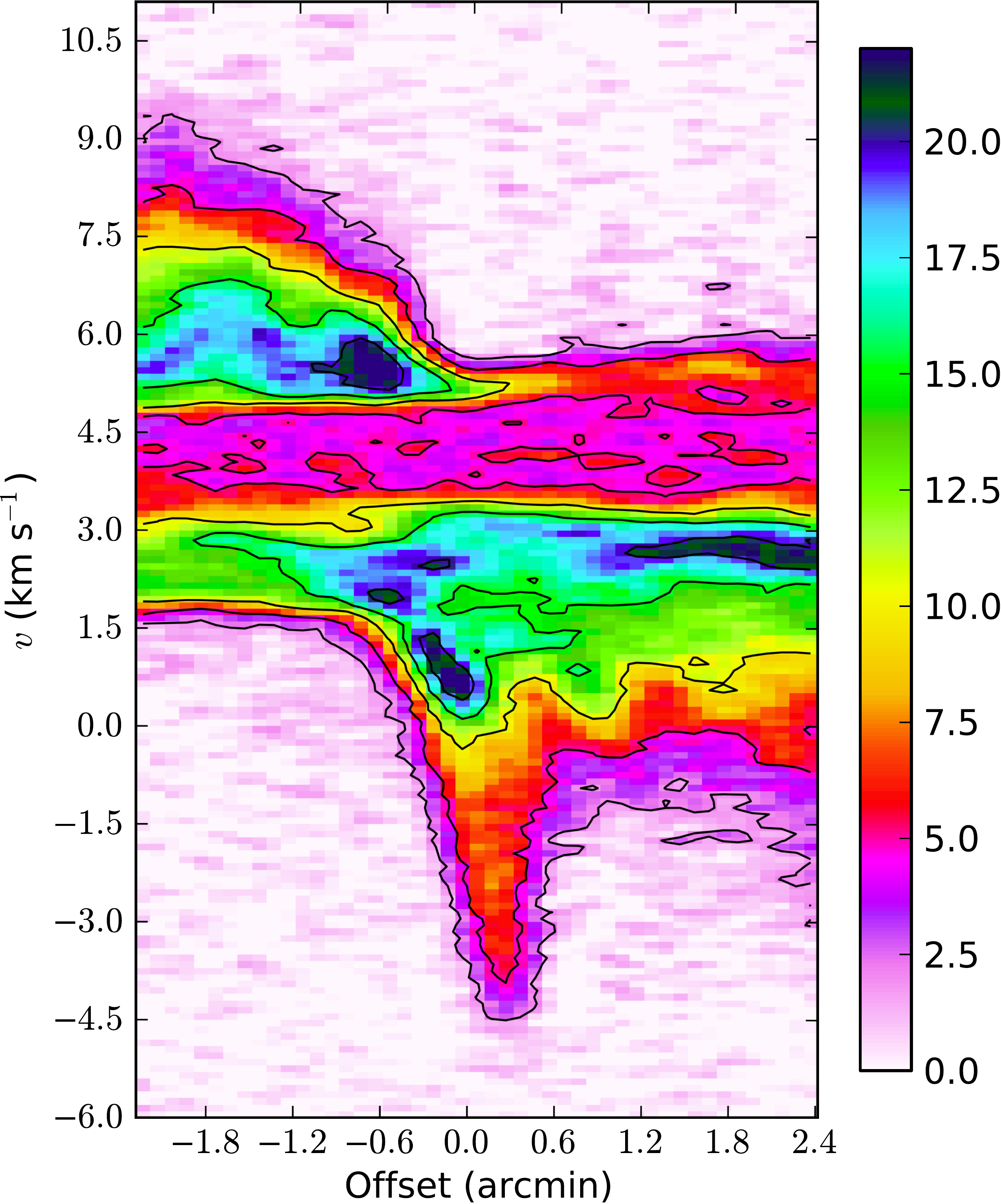}
\includegraphics[width=3in]{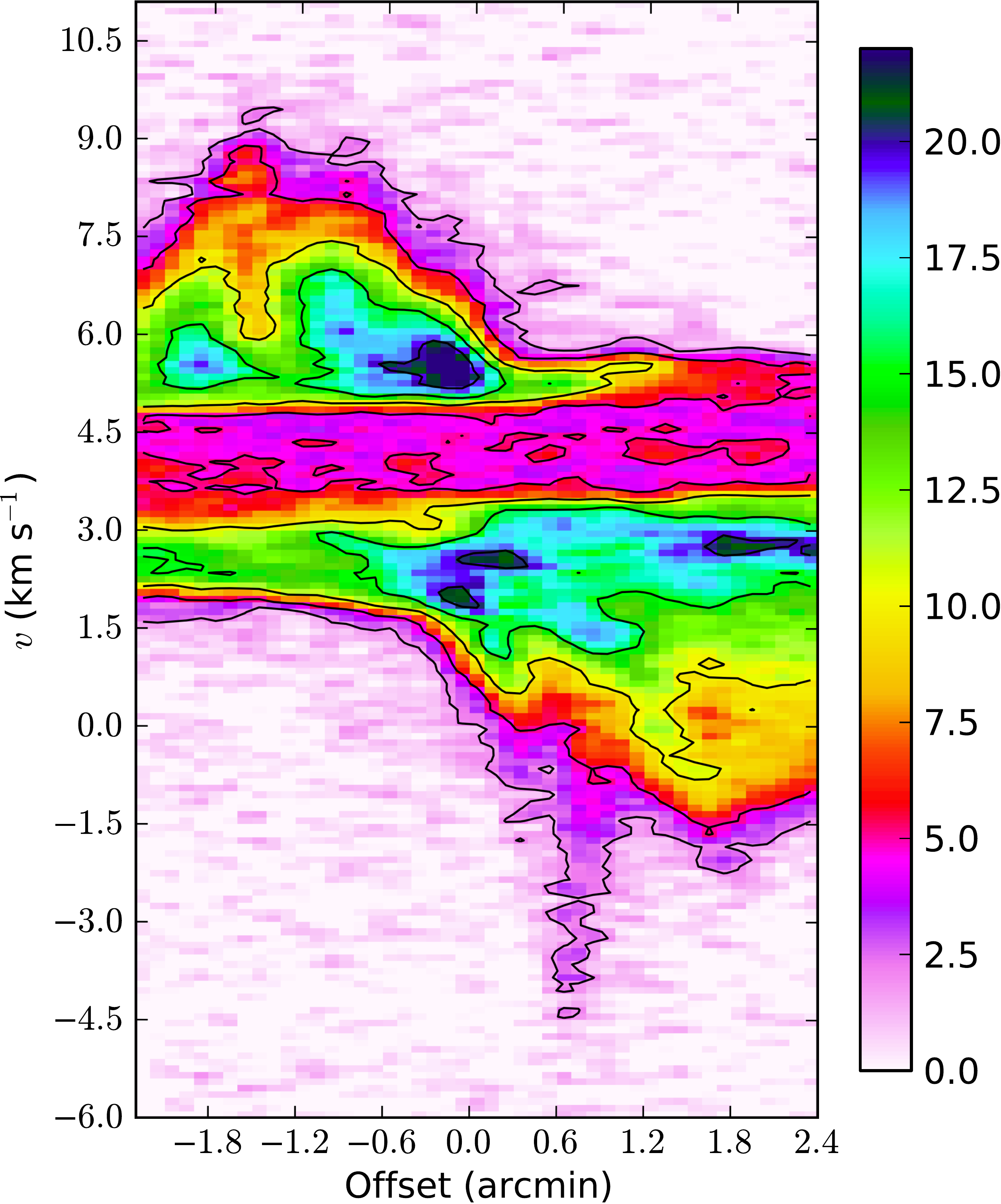}
\includegraphics[width=4in]{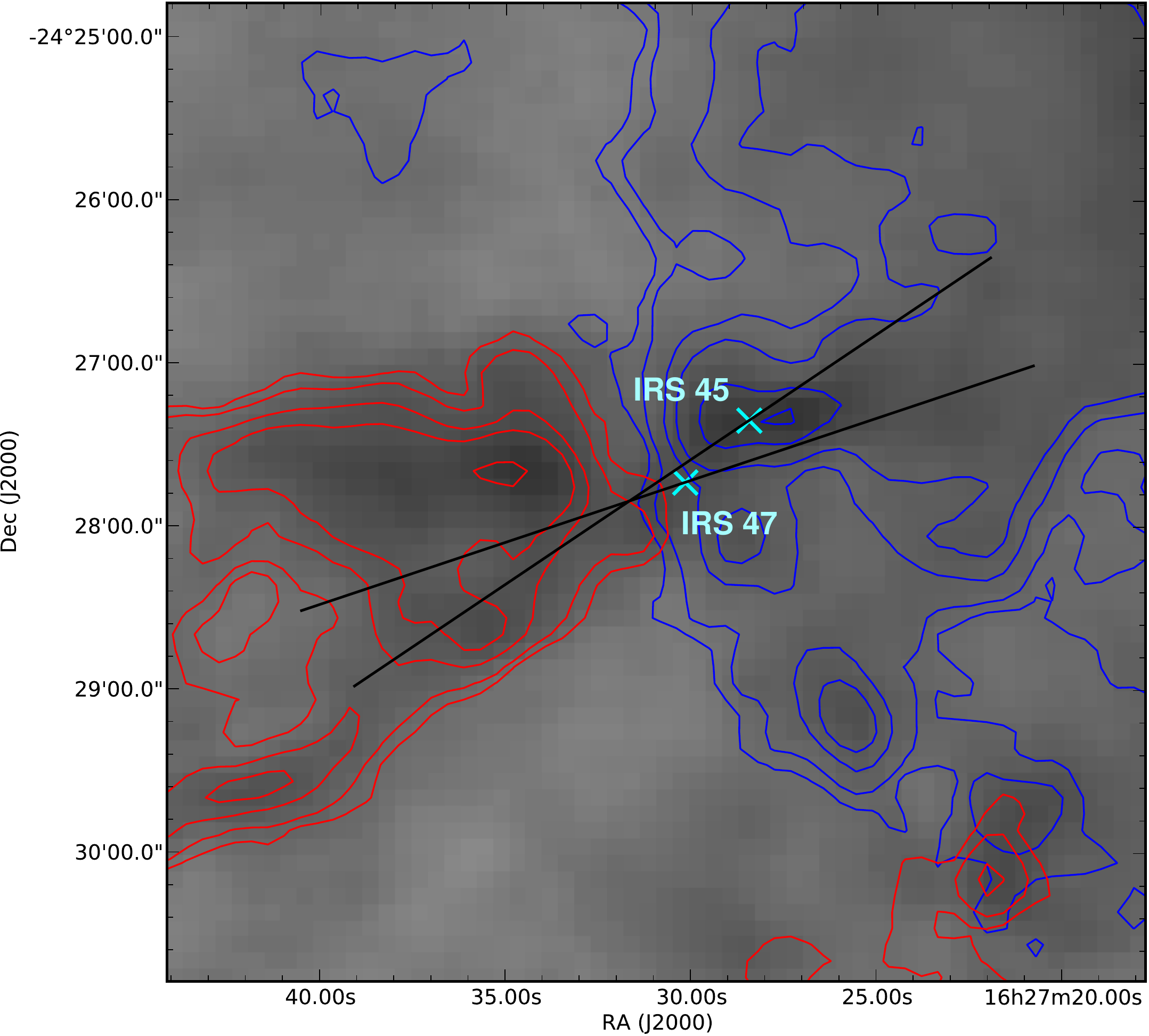}
\caption{Position-velocity diagrams for sources IRS~45 (left) and IRS~47 (right).  The cuts used for the position-velocity (PV) diagram are shown in the lower panel, along with the red and blue outflow lobes. The colour bar is shown in units of main beam brightness temperature~T$_{MB}$.}
\label{fig:OphB_pv}
\end{figure*}

Protostars SSTc2d~J162730.9-242733, IRS~45, and IRS~47 can be seen in the central region of a large, clumpy outflow in Oph~B (Figure~\ref{fig:OphB_br}).  Kamazaki \al (2001) attributed this outflow to IRS~45 due to its proximity to the strongest blueshifted emission in the lobe.  The position-velocity diagram in Figure~\ref{fig:OphB_pv} clearly shows the redshifted lobe extending from the centre of IRS~47 and $\sim$0.3$^{\prime}$ from IRS~45, also seen in Figure~\ref{fig:OphB_br}.  We therefore suggest that IRS~47 is the driving source for this region due to its more central location between the blue and redshifted lobes, in agreement with the suggestion by van der Marel \al (2013).

The bipolar outflow driven by Class~I source WL~6 is located south-west of IRS~47.  Khanzadayan \al (2004) has suggested that this source may drive the H$_2$~knot f09-02 from the blueshifted lobe of the outflow.  This outflow detection agrees with the observations of Sekimoto \al (1997).  The last firm bipolar outflow detection is from IRS~54 in the east of Oph~B.  IRS~54 is a binary with a 7\arcsec separation based on near-IR observations (Haisch \al 2006, Duch$\hat{e}$ne \al 2004), which has been suggested by J$\o$rgensen \al (2009) to drive a bipolar precessing outflow.  This source is also associated with H$_2$ knots [G01]~C1--5 (Grosso \al 2001) and f08-01~a--h (Khanzadayan \al 2004, G$\acute{\rm{o}}$mez \al 2003).  In Figure~\ref{fig:OphB_br}, the location of the H$_2$ knots and the high-velocity molecular gas associated with the outflow trace a curved path, further supporting the presence of a precessing outflow, similar to that reported in L1551 by Fridlund \al 1989, 2002, 2005) and White \al (2006). 

Sources IRS~37 and SSTc2d~J162721.8-242728 also have marginal molecular outflow detections with the blueshifted high velocity lobe present.  These sources are located near the blueshifted lobe of the IRS~47 outflow, which is potentially the cause of the blue line-wing detection.  No detection of an outflow was found for SSTc2d~J162716.4-243114, a flat spectrum source south of the main Oph~B outflow.

\subsection{Oph~C and E}
\label{OphCE}

\begin{figure}
\centering
\includegraphics[width=3.35in]{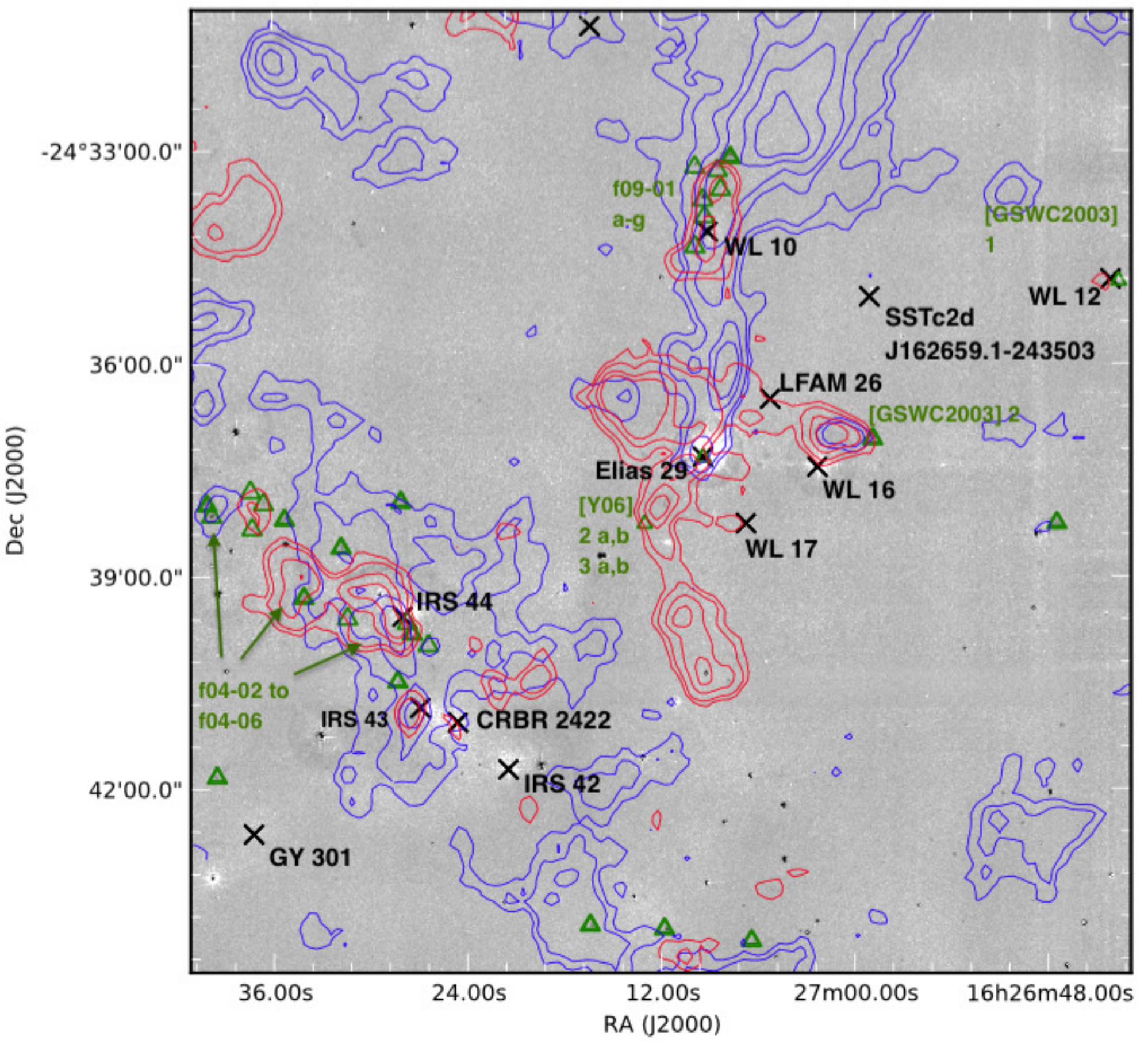}
\caption{CO~${\it J}$ = 3-2 outflows in Oph~C, E, and F. Blue contours trace the blueshifted CO intensity $\int T_{\mathrm{MB}} \ {\mathrm{dv}}$ (integrated from $-5.7$ to $1.3$~km~s$^{-1}$) and red contours trace the redshifted CO intensity (integrated from 7.0 to 14.3~km~s$^{-1}$; contours are at 3, 5, 10, 15, 30, and 45~K~km~s$^{-1}$}
\label{fig:OphCEF_br}
\end{figure}

The Oph~C and E regions lie in the western/south-western of LDN1688. Although Oph-C lies within the area mapped by the CO observations, it is located just at the edge of the $^{13}$CO map, and just off edge of the C$^{18}$O map shown in Figure \ref{Fig:c18o_cores}.  Together, these regions contain six Class~I protostars and one Class~II protostar.  Using the CO outflow criteria, two sources (Elias~29 and WL~10) are confirmed to drive outflows and five sources are identified as marginal molecular outflow candidates, with one having a high velocity line-wing detected.   Figure~\ref{fig:OphCEF_br} shows the H$_2$ image with the blue and redshifted CO outflows for Oph~C, E, and F superimposed as blue and red contours respectively.

The bipolar outflow driven by Elias~29 can be clearly seen in Figure~\ref{fig:OphCEF_br}.   This outflow is known for its s-shape (Bontemps \al 1996, Sekimoto \al 1997, Bussmann \al 2007), which has been suggested to be a consequence of a precessing outflow associated with a binary system (Ybarra \al 2006), similar to that of L1551 (White \al 2006). The outflow encompasses several H$_2$ knots, mainly in the southern redshifted lobe, which are labelled 2~a,b and 3~a,b from Ybarra \al 2006. G$\acute{\rm{o}}$mez \al (2003) also suggest that the knot [GSWC2003]2 is driven by Elias~29, aligning with the east-west CO emission that appears to be driven by the source.  The blueshifted outflow lobe extends to the north, also potentially driving several knots near the Class~II protostar WL~10.  WL~10 lies north of Elias~29 in the Oph~C region and is known to be associated with a bipolar outflow (Sekimoto \al 1997) that can also be seen in Figure~\ref{fig:OphCEF_br}.  The outflow encompasses H$_2$ knots f09-01~a--g (Khanzadayan \al 2004). G$\acute{\rm{o}}$mez \al (2003) and Zhang \al (2011) have attributed f09-01~b,e--g to the outflow, although f09-01~a,c,d may also be related to the CO outflow from WL~10.  The latter knots have been previously been attributed to WL~16 (G$\acute{\rm{o}}$mez \al 2003, Zhang \al 2011).

Sources LFAM~26, WL~17, and WL~16 lie near Elias~29 and could suffer confusion from the high-velocity emission from its outflow.  LFAM~26 and WL~17 have both blue and redshifted line-wings, and LFAM~26 has been classified as a driving source for an outflow by previous studies (Bussmann \al 2007, Nakamura \al 2011).  Neither of these sources have strong bipolar lobes in Figure~\ref{fig:OphCEF_br} and both lie directly to the west of Elias~29.  WL~16 has a faint detection of a blue line-wing and past studies (G$\acute{\rm{o}}$mez \al 2003, Zhang \al 2011) have suggested that the source drives H$_2$ knots f09-01~a,c,d near WL~10.  WL~12 is located in the western portion of Oph~C and has only a slight redshifted line-wing  Bontemps \al (2006) classified this source as driving a molecular outflow but considered it a more uncertain detection because the source did not have a bipolar detection, which is reflected in our study.  Lastly, Class~I source SSTc2d~J162659.1-243503 has a faint detection of a red line-wing.  

\subsection{Oph~F}
\label{OphF}

The Oph~F region makes up the south-eastern portion of the LDN1688 cloud, and was covered in all three of the isotopologues shown in Figure \ref{Fig:c18o_cores}.  The region contains four Class~I and one flat spectrum protostars, each of which was examined for the outflow study.  Using the CO outflow criteria, two sources (IRS~43 and IRS~44) are confirmed to drive outflows, two sources are identified as molecular outflow candidates with one high velocity line-wing detected, and one source does not have a clearly discernible associated outflow. In Figure~\ref{fig:OphCEF_br}, the bipolar outflows driven by IRS~43 and IRS~44 can be clearly seen.  These outflow sources have been confirmed by past studies (IRS~43:  Bontemps $\&$ Andr$\acute{e}$ 1997; IRS~44: Bontemps \al 1996, Bussmann \al 2007).  H$_2$ knots attributed to the two sources are labelled f04-02 to f04-06 (Khanzadayan \al 2004).  IRS~42 and CRBR~2422 are located south of IRS~44 and IRS~43.

There were marginal outflow detections from the sources CRBR~2422 and IRS~42, located to the south-west of IRS~43 and IRS~44.  These sources only have a faint red line-wing detected.  Due to their proximity to IRS~43 and IRS~44, the line-wing detection is most likely the result of confusion caused from the nearby outflows.  Additionally, there was not an outflow detection from flat spectrum source GY~301.  

\subsection{Outflow mass and energetics}
\label{global}

\begin{figure*}
\includegraphics[width=7in]{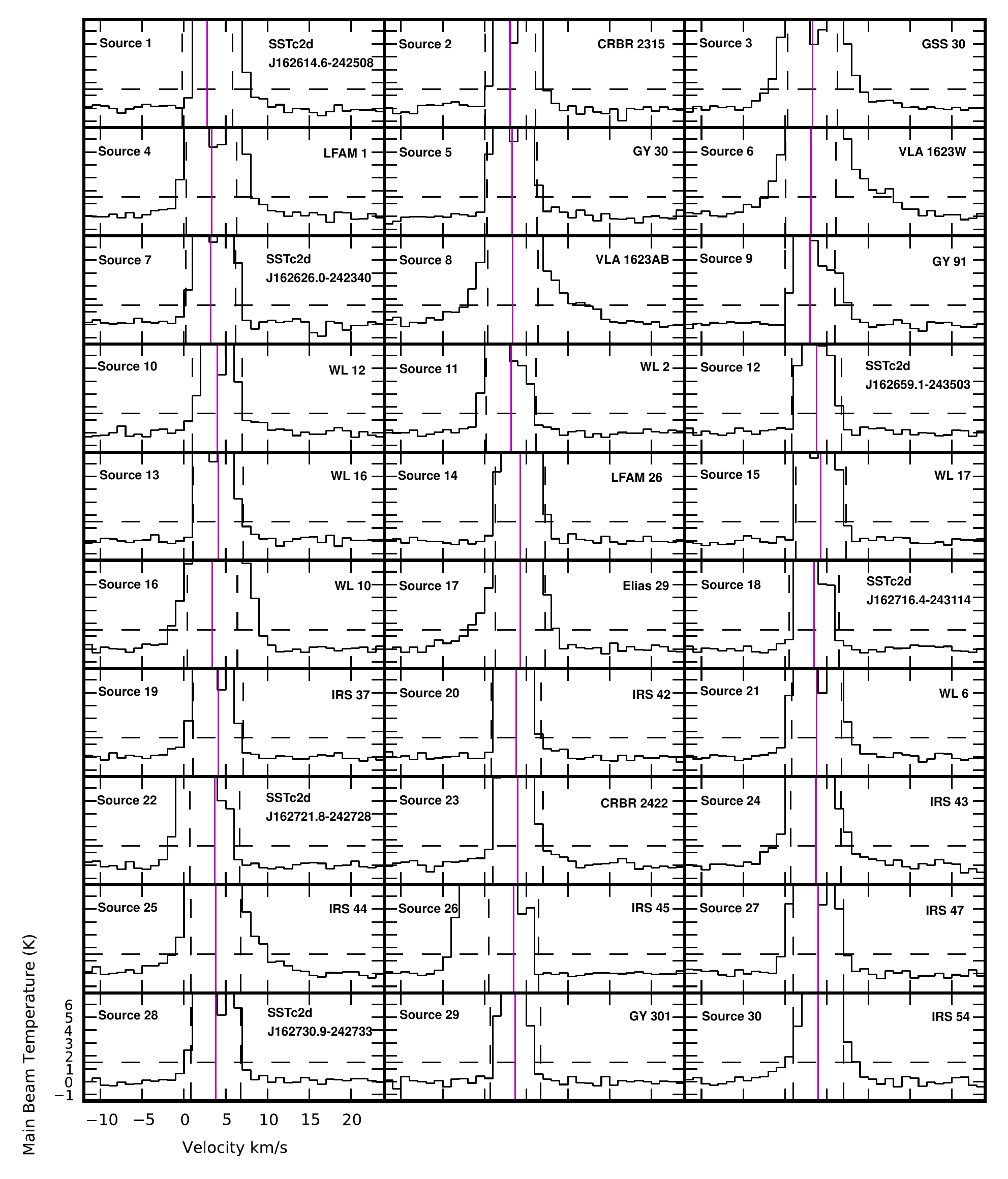}
\caption{CO spectra of sources listed in Table \ref{source_list}, where the solid vertical line represents the systemic velocity (i.e. v$_{LSR}$ of the line centroid of the C$^{18}$O line at the same position, the dashed vertical lines are drawn at $\pm$3 \kmsec~from this, and the dashed horizontal lines show the \Tmb = 1.5 K level).}
\label{fig:12co}
\end{figure*}

The CO~${\it J}$ = 3--2 gas at high blue and redshifted velocities relative to the systemic velocity, v$_{LSR}$ (e.g. see Figure \ref{fig:12co}), was used to calculate the global outflow mass and energetics of the cloud.  To exclude the contribution from ambient cloud emission, the central line velocity, $\mathrm{v_0}$, was estimated by using the average C$^{18}$O spectrum of the cloud, which was then fitted with a Gaussian to determine the line centre, and found to be 3.3~km~s$^{-1}$. The line criterion used to identify the presence of outflows was (following Hatchell \al 2007) to select spectra where the linewidths were broader than that assumed to be associated with ambient gas, by searching for lines where \Tmb at offsets of $\pm$ 2.5~km~s$^{-1}$ from $\mathrm{v_0}$~$>$ 1.5 K. 

The criterion used to estimate the global energetics is similar to the outflow criterion but it takes into account global effects including the changes of the line centre of up to 1~km~s$^{-1}$ seen across the cloud, so as to better estimate the velocity range counted as ambient over that of the outflow gas.  However, some ambient emission will be included in these calculations because there is no fixed velocity boundary between outflow-driven gas, and that of gas set in motion by ambient cloud turbulence. CO spectra of sources listed in Table \ref{source_list} are shown in Figure \ref{fig:12co}. Specifically the central source velocities were calculated by fitting a gaussian to each individual C$^{18}$O spectrum, which should trace the ambient cloud rather than the outflow gas.

\subsubsection{Line opacities in the outflow line wings}
\label{opacity}

The CO and $^{13}$CO \Tmb ratio can be used to estimate the CO opacity per velocity channel, $\tau_{12}(\mathrm{v})$.  The maps were averaged to a single spectrum and clipped at 3$\sigma_{\mathrm{rms}}$ and 1$\sigma_\mathrm{rms}$ for CO and $^{13}$CO respectively.  This enhanced the velocity range ($\sim$~0--9~km~s$^{-1}$) allowing $^{13}$CO to be detected in the line-wings for the opacity calculation. The CO optical depth was then estimated using the following relation (e.g. Hatchell \al 1999b, Curtis \al 2010, Graves \al 2010):

\begin{equation}  \frac{T_\mathrm{12}}{T_\mathrm{13}} = \left( \frac{\nu_{\mathrm{12}}}{\nu_{\mathrm{13}}} \right)^2  f_{13}  \left( \frac{1 - \exp ({-\tau_{\mathrm{12}}})}{\tau_\mathrm{12}} \right )
\label{Eqn:CO_opacity}
\end{equation}

where $T$ is the main-beam temperature of each CO transition, $\nu$ is the frequency of the isotopologues, $f_{13}$ is the relative abundance of CO to $^{13}$CO (77; Wilson $\&$ Rood 1994), and $\tau_\mathrm{12}$ is the optical depth of CO. This assumes that $^{13}$CO is optically thin and CO is optically thick (see Hatchell \al 1999a, Curtis \al 2010). In this case, the ratio of observed antenna temperatures, T(CO)/T($^{13}$CO) = (1-exp(-$\tau_{12}$))/(1-exp(-${\tau_{13}}$)). where ${\tau_{12}}$ = X ${\tau_{13}}$, where X is the abundance ratio between the two molecules. Then, if ${\tau_{13}}$ is optically thin, 1-exp(-${\tau_{13}}$) = ${\tau_{13}}$. This approach accounts for the differences in frequency of occurrence between the two molecules, with the ratio between the temperature of the two molecules decreasing with increasing CO optical depth.

Even though CO is expected to be optically thin in the high velocity blue and redshifted line-wings, $^{13}$CO does not extend as far in velocity range as the CO emission.  Therefore, this equation was solved by assuming $\tau_\mathrm{12} \rightarrow \infty$ so that $1 - \exp (-\tau_\mathrm{12}) \approx 1$.

\begin{figure}
\centering
\includegraphics[width=3.4in]{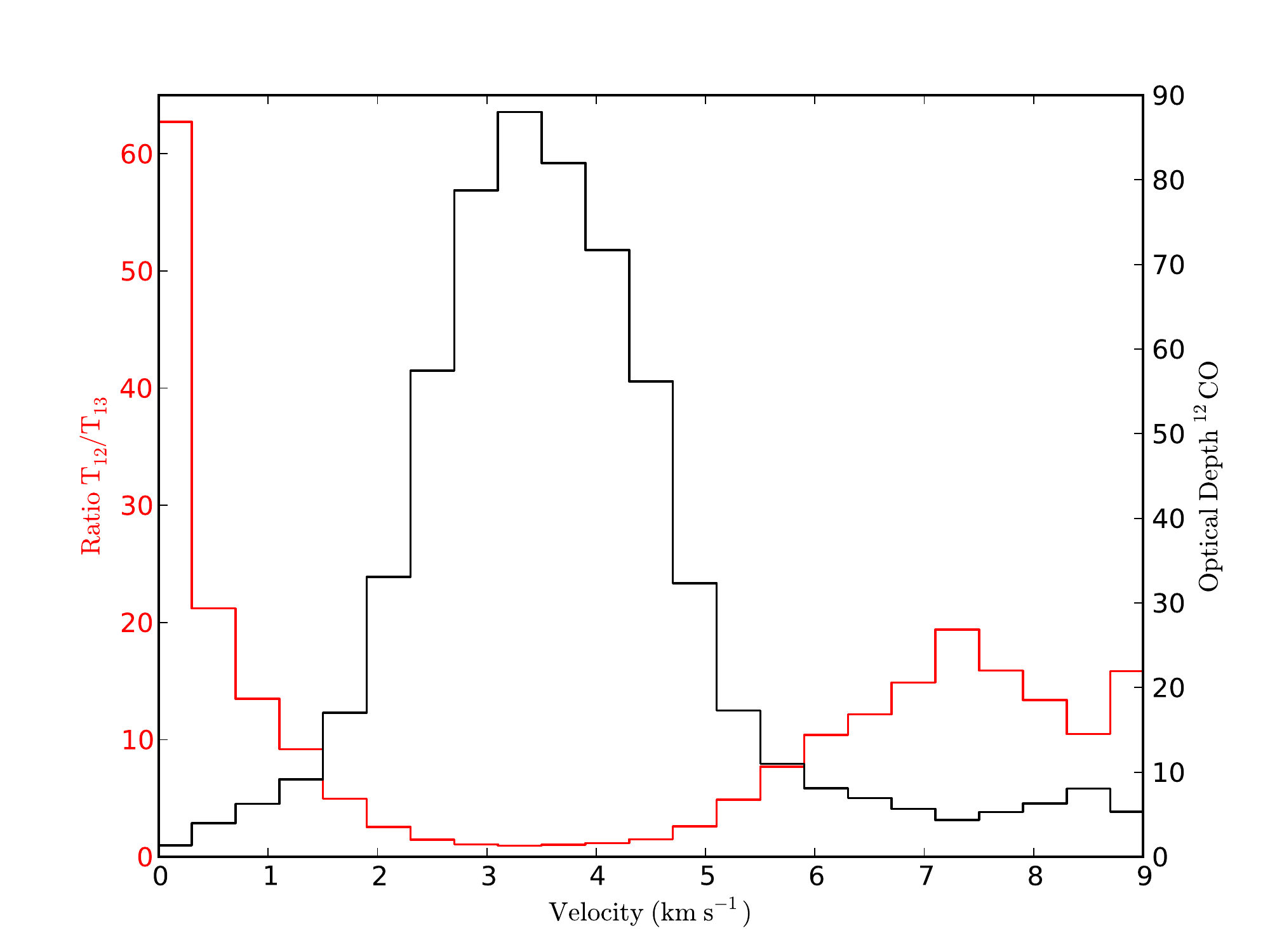}
\caption{CO optical depth variation with velocity, $\tau_\mathrm{12}(\mathrm{v})$.  The ratio of CO to $^{13}$CO is shown as a function of velocity (red) and the CO optical depth is shown for the same velocity region (black). }
\label{fig:opacity}
\end{figure}

Figure~\ref{fig:opacity} shows the variation of the ratio of CO to $^{13}$CO main-beam temperatures with the calculated CO optical depth per velocity channel.  The CO emission is optically thick ($\tau_\mathrm{12} > 1$) over the velocity range 0 $\rightarrow$~9~km~s$^{-1}$. The limit of the CO optical depth measurement is $\tau_\mathrm{12} \sim 3$.  The ratio of CO and $^{13}$CO main-beam temperatures supports the suggestion that CO is optically thick because the ratio only begins to approach the abundance ratio at the edges of the velocity range.  We note the ratio at redshifted velocities ($\sim$~7--9~km~s$^{-1}$) does not fully reach the abundance ratio, indicating that CO is still optically thick in this range. The opacity correction applied to the CO main-beam temperature is further discussed in Section~\ref{global}.

Assuming local thermodynamic equilibrium, the outflow masses were calculated assuming a distance of 120~pc, with and without optical depth corrections per velocity channel, and an assumed excitation temperature of 50~K (Hatchell \al 2007, Curtis \al 2010, Graves \al 2010).  The column density of a diatomic molecule may be estimated using the  emission from a line (such as the $(J+1)$ $\rightarrow$ $J$ transition) of an optically thin(ner, than CO) isotopomer; for the example of the ${\it J}$ = 3--2 transition (following Richardson \al 1985, Kamazaki \al 2003), and assuming local thermodynamic equilibrium, the mass can be estimated as:

\begin{eqnarray}
M_{\mathrm{CO}} &=& 3.77 \ \times \ 10^{-8} \left({\frac{X_{\mathrm{CO}}}{10^{-4}}}\right)^{-1} \left({\frac{\mathrm{d}}{120 \ \mathrm{pc}}}\right)^2   \\ \nonumber &\times& \left({\frac{A_{\mathrm{pixel}}}{\mathrm{arcsec}^2}}\right) \  \displaystyle\sum\limits_{j} \left(\int{T_{\mathrm{MB},j}} \ \mathrm{dv} \right) \ M_\odot
\label{Eqn:mass}
\end{eqnarray}

where $X_{\mathrm{CO}}$ is the relative abundance of CO to H$_2$ (Blake \al 1987), $A_{\mathrm{pixel}}$ is the pixel area (36~arcsec$^2$), d is the distance (in parsecs), and $j$ is an index over map pixels.

To estimate the momentum and energy, $\int T_\mathrm{MB} \ \mathrm{dv}$ is replaced by $\int T_{\mathrm{MB}} |\mathrm{v - v_0}| \ \mathrm{dv}$ and $\frac{1}{2} \int T_\mathrm{MB} (\mathrm{v - v_0})^2 \ \mathrm{dv}$ respectively, where $\mathrm{v_0}$ is the line centre for the cloud.  With the correction for CO optical depth, the integrated-main beam intensity becomes
\begin{eqnarray} \int T_\mathrm{corr} \ \mathrm{dv} = \int T_\mathrm{MB} \ \frac{\tau_\mathrm{12}}{1 - \exp{(- \tau_\mathrm{12})}} \ \mathrm{dv}
\end{eqnarray}

and the momentum and energy estimates are corrected similarly.  Blue- and redshifted main-beam temperatures are integrated from -8~$\rightarrow$~0.8 and 5.8~$\rightarrow$~15~km~s$^{-1}$ respectively and the opacity correction is applied to the corresponding velocity channels between 0~$\rightarrow$~-9~km~s$^{-1}$.  The velocity channel limits were based on whether or not there was detectable CO emission at the 1$\sigma_\mathrm{rms}$ level.  
\begin{table}
\centering
\begin{tabular}{l c c c}
\hline
& Mass & Momentum & Energy \\
& ($M_\odot$) & ($M_\odot$~km~s$^{-1}$) & ($M_\odot$~km$^2$~s$^{-2}$) \\ \hline
Thin & & & \\
  Blue & 0.20 & 0.74 & 1.49  \\
  Red & 0.54 & 1.74 & 3.18 \\
Corr. & & & \\   
  Blue & 0.42 & 1.36 & 2.36 \\
  Red & 4.30 & 12.67 & 19.85 \\ \hline
\end{tabular}
\caption{Global kinematics for the out-flowing gas assuming optically thin CO line-wing emission (thin) and with an optical depth correction (corr.).}
\label{global_properties}
\end{table}

The basic kinematics for the global out-flowing gas in Ophiuchus are summarised in Table~\ref{global_properties}.  The mass, momentum, and energy increase substantially by factors of 6.4, 5.7, and 4.8 respectively when the optical depth of CO is taken into account.  Effects involving outflow inclination angle can also greatly reduce the amount of momentum and energy observed.  A correction for inclination can be applied to the blue and redshifted velocities for momentum and energy calculations.  The line of sight velocities should be corrected by a factor of $1 / \cos(i)$, where $i$ is the inclination angle.  Assuming random inclinations to the line of sight with uniformly distributed outflows ($i \approx 57.3\ \mathrm{deg}$; Bontemps \al 1996), the momentum will increase further by a factor of 2 and the energy by a factor of 3 (Curtis \al 2010). A detailed discussion of the efficacy of outflow properties as estimated from molecular line data is given by van der Marel \al (2013), who show that the main techniques adopted in most studies agree to within a factor of $\sim$6--10, differences of which can be reconciled by comparing the analysis methods. 

The kinetic energy generated by the global high-velocity outflows can also be compared to the overall turbulent kinetic energy.   When inclination and optical depth are taken into account, the total outflow kinetic energy is $1.3 \times 10^{38}$~J (67~$M_\odot$~km$^2$~s$^{-2}$) or 21\% of the turbulent kinetic energy shown in Table~\ref{energy}.  This suggests that outflows are significant but not the dominant source of turbulence in Ophiuchus.  Conversely, Nakamura \al (2011) suggested that protostellar outflows can drive turbulence in the region after finding the total outflow energy injection rate ($L_\mathrm{tot} \sim 0.2~L_\odot$) larger than the turbulent dissipation rate of the supersonic turbulence ($L_\mathrm{turb} \sim 0.06$--$0.12~L_\odot$) assuming optically thin emission.  A similar method was used to calculate the global outflow energy injection rate $L_\mathrm{global} = E_\mathrm{out} / T_\mathrm{I}$ where $E_\mathrm{out}$ is the global outflow kinetic energy and $T_\mathrm{I}$ is the lifetime of the Class~I protostellar phase ($\sim$~0.5~Myr; Evans \al 2009).  This gives a global outflow energy injection rate $L_\mathrm{global} = 0.02$~$L_\odot$, which is a factor of 10 smaller than the injection rate from Nakamura \al (2011).  The difference between the outflow energy injection rates is mainly due to our assumption that a longer outflow time-scale better reflects the age of Class~I protostars, as opposed to calculating the dynamical time-scales of individual outflows.
  
Comparing our lower outflow injection rate to the turbulent dissipation rate from Nakamura \al (2011) suggests that the global outflow injection rate is lower than the turbulent dissipation rate by a factor of 3--6.  Additionally, optical depth corrections can be made to the turbulent dissipation rate which would further increase the difference between the injection and dissipation rates.  This result agrees with our conclusion that global outflows in the Ophiuchus region are significant but not necessarily the dominant driver of turbulence in the cloud.

Supersonic turbulence would however dissipate rapidly, requiring that a power source other than outflows would have to be sustained for Ophiuchus. The energy needed might potentially come as a consequence of virialisation as clouds/clumps collapse, with turbulence generation occurring on a time scale of a few dynamical times ($\sim$10$^6$ yr for gas with a density $\sim$ 10$^4$ cm$^{-3}$ from Hocuk $\&$ Spaans 2010).

The mass of the cloud estimated earlier in Section \ref{cloud_energetics:opacities} using the C$^{18}$O integrated intensity, and the virial mass of the cloud assuming a spherical cloud of uniform density (Rohlfs $\&$ Wilson 2000) are both estimated and summarised in Table~\ref{energy}.

\begin{table*}
\begin{minipage}{7in}
\centering
\begin{tabular}{l c c c c}
\hline
Cloud & Total mass & ${\it E}_\mathrm{kin}$ & $E_\mathrm{grav}$ & Outflow Energy \\
& ($M_\odot$) & (J) & (J) & (J) \\
\hline
Ophiuchus & 515 & $6.3 \times 10^{38}$ & $4.5 \times 10^{39}$ &  $1.3 \times 10^{38}$ \\
\hline
\end{tabular}
\caption{Global mass and energetics.}
\label{energy}
\end{minipage}
\end{table*}

For Ophiuchus, the gravitational binding energy is estimated to be  $4.5 \times 10^{39}$ ~J (2282~$M_\odot$~km$^2$~s$^{-2}$), and by using $\mathrm{v_{fwhm}}$ from the average C$^{18}$O linewidth across the cloud, the turbulent kinetic energy is estimated to be $6.3 \times 10^{38}$~J (320~$M_\odot$~km$^2$~s$^{-2}$).  The turbulent kinetic energy is therefore roughly a factor of 7 smaller than the gravitational binding energy, suggesting that the cloud is gravitationally bound.

\section{Molecular clump identification and analysis}
\label{clump_analysis}

The basic properties of the spatial distribution of the gas in the Ophiuchus complex were estimated using the $\it{CPROPS}$ package (Rosolowsky $\&$ Leroy 2006) to identify clumplets, based on the C$^{18}$O maps (because this isotopomer is the least opaque of those studied). The properties of molecular clumplets were then estimated along with information from the other isotopomeric lines, that minimised the biases introduced by finite signal-to-noise and spatial resolution. The recovered data were then used (by assuming that all material was at the adopted distance to the Oph cloud) to estimate the molecular line luminosities and other properties such as their temperatures.

$\it{CPROPS}$ initially searches for contiguous, significant emission ($\ge$ 3$\sigma$) compared to nearby emission free parts of the spectrum) in the C$^{18}$O position-velocity data cube, after decomposing the area surveyed into individual clumplets. Regions having areas less than that of two beam footprints are discarded, along with any detections where the emission was only detected in a single (0.1 km s$^{-1}$) velocity channel. To account for diffuse emission from the turbulent bulk of the cloud, the C$^{18}$O data were filtered using an {\it \`atrous} wavelet algorithm, similar to that described in Alves, Lombardi $\&$ Lada (2007).  This algorithm removes large scale emission, whilst preserving small scale clump structures, without introducing significant negative features, which is a common limitation of Fourier filtering. This also helps to avoid the confusion imparted by the background emission, which has in the past been another limiting factor in estimation of clump mass spectra when using algorithmic approaches. Specifically, we removed scales on the top three levels of a $B_3$ cubic spline wavelet transform (see Starck $\&$ Murtagh 2006 for details), that corresponded to scales of 0.2 pc and larger.  We then used the $\it{CPROPS}$ algorithm to identify clumplets within the filtered cube, identify local maxima in the data, find shared surfaces, and to estimate the relative proportions of the emission attributable to each of the cloud clumplets. The algorithm identifies regions of position-position-velocity space associated with clumplets to calculate the properties of clumplets in the original C$^{18}$O data (pre-filtering), as well as of the corresponding regions in the $^{13}$CO and CO data sets.  This joint analysis produces a matched catalogue of clumplets across all three isotopologue data cubes.

The default implementation of this is a watershed algorithm, similar to the widely used Clumpfind program, but which does not `fight' over emission that is contested between objects. A first order correction for sensitivity bias was made using the methodology described by Rosolowsky $\&$ Leroy (2006) to extrapolate the measured parameters of the cloud structures to those expected for a cloud(s) with a boundary iso-surface of T$_{edge}$ = 0 K (i.e. assuming perfect sensitivity and an isolated single source, following the original formulation (Blitz $\&$ Thaddeus (1980), Scoville \al 1987). Finally, the properties of the individual clumplets were measured, based on the extracted values from $\it{CPROPS}$. The performance of this clump extraction technique has been shown to compare favourably with other extraction codes such as $\it{GAUSSCLUMPS}$ and $\it{CLFIND}$ (see Rosolowsky $\&$ Leroy 2006, and Bolatto \al 2008 for details). The robustness, reliability, and consistency of $\it{CPROPS}$ appears to be very good for clouds having signal-to-noise ratios (s/n)~$>$~10.

The clump properties were estimated using moments of the intensity distribution.  The radii (${\it R}$) and velocity dispersions ($\sigma_v$) were calculated from the intensity-weighted second moments in the spatial and velocity dimensions respectively, with the values of the radii reported in this paper being deconvolved from the gaussian telescope beamwidth.  Clump masses were then characterised by a virial and a luminous mass estimate. The virial mass estimate is calculated adopting a $\rho(r) \propto r^{-1}$ profile, following Scoville et al. (1987). Luminous masses were then estimated from the integrated CO flux $S_{\mathrm{CO}}$ using a constant CO-to-H$_2$ conversion factor:

\begin{equation}
M_{lum} = 0.011 S_{\mathrm{CO}} \left[\frac{X_{\mathrm{CO}}}{2\times 10^{20} \mathrm{~cm^{-2}~(K~km~s^{-1})^{-1}}}\right] D^2 M_{\odot}  
 \label{Eqn:m_lum}
\end{equation}

where $S_{\rm CO} = \int{I_{\rm CO}}~d\Omega~dv$ is the integrated CO flux in units of Jy km s$^{-1}$, $D$ is the distance to the source in kpc, and $X_{\mathrm{CO}}$ is the CO-to-H$_2$ conversion factor appropriate for CO$(1-0)$ data, and includes a contribution to the mass from He (see Bolatto et al. 2008 for further details). It should be noted that $X_{\mathrm{CO}}$ may vary for very high CO ${\it J}$-transitions, if these are dominated by higher excitation material (see detailed discussion of this by Papadopoulos \al 2012a,b), but at the moment there is no evidence to suggest a significant variation to this quantity between the CO ${\it J}$ = 1--0 and 3--2 transitions. As a further cautionary note, that the Ophiuchus region suffers from widespread self-absorption by line of sight CO, and further we note that the C$^{18}$O line is optically thick at a few locations - both of these effects may in practice introduce some uncertainty to the luminous, and virial mass estimates (see Rosolowsky $\&$ Leroy 2008 for further details), although there is no reason to believe that these will provide a significant bias to the results.

The results of the clump decomposition are shown in Table \ref{Table:clumps}, with the full catalogue of 105 clumplets being given in the supplementary material to this paper.

\begin{table*}
\vspace{0pt}
\caption{The clumplet catalogue (a full version is available in the Supplementary material on-line). The clumplets listed in the catalogue are: (1) a short form running number, (2,3) Right Ascension and Declination (J2000) of the clump centre, (4,5,6) the major and minor axis of the clump deconvolved from the beam in arc seconds, along with the position angle (measured east of north), (7) the deconvolved radius of the clump as defined by Rosolowsky \al (2006), (8,9) centre v$_{LSR}$ and dispersion of the clump as determined from the C$^{18}$O data, (10) The virial mass, and (11) the Luminous mass estimated from the CO observations.}
\begin{scriptsize}
\fontsize{8}{10}\selectfont
\begin{tabular}{l l l r r c c c c c c r r}
\hline
\multicolumn{1}{l}{No} & \multicolumn{1}{c}{RA (J2000)} & \multicolumn{1}{c}{Dec (J2000)} & \multicolumn{1}{c}{Maj} & \multicolumn{1}{c}{Min} & \multicolumn{1}{c}{Pos angle} & \multicolumn{1}{c}{Radius} & \multicolumn{1}{c}{Velocity} & \multicolumn{1}{c}{Dispersion} & \multicolumn{1}{c}{Virial mass} & \multicolumn{1}{c}{Luminosity} \\ 
 \multicolumn{1}{l}{} & \multicolumn{1}{c}{h:m:s.s}   & \multicolumn{1}{c}{d:m:s.s} &  \multicolumn{1}{c}{${\prime\prime}$} & \multicolumn{1}{c}{${\prime\prime}$} &  & \multicolumn{1}{c}{pc} & \multicolumn{1}{c}{km s$^{-1}$} & \multicolumn{1}{c}{km s$^{-1}$}  & \multicolumn{1}{c}{\Msun}  & \multicolumn{1}{c}{K km s$^{-1}$ pc$^{-2}$} \\
\multicolumn{1}{l}{(1)} & \multicolumn{1}{c}{(2)} & \multicolumn{1}{c}{(3)} & \multicolumn{1}{c}{(4)} &  \multicolumn{1}{c}{(5)} & \multicolumn{1}{c}{(6)} & \multicolumn{1}{c}{(7)} & \multicolumn{1}{c}{(8)} & \multicolumn{1}{c}{(9)} & \multicolumn{1}{c}{(10)} & \multicolumn{1}{c}{(11)}  \\
\hline
1&16:25:54.0&-24:29:46.7&22.9&16.1&51&0.02&2.95&0.20&0.71&0.00141\\
2&16:25:55.4&-24:17:30.2&7.9&7.7&-38&0.00&3.54&0.11&0.04&0.00015\\
3&16:25:56.9&-24:23:25.3&12.8&11.2&-41&0.01&3.45&0.17&0.27&0.00034\\
4&16:26:00.1&-24:31:24.7&42.1&13.2&2&0.02&3.41&0.28&1.62&0.00147\\
5&16:26:00.4&-24:33:53.0&8.4&8.0&63&0.00&3.69&0.17&0.13&0.00022\\
6&16:26:01.3&-24:32:29.1&26.6&14.2&132&0.02&3.35&0.26&1.18&0.00146\\
7&16:26:02.1&-24:18:45.2&11.0&27.2&65&0.01&3.75&0.32&1.51&0.00781\\
8&16:26:02.5&-24:25:31.2&13.6&12.4&118&0.01&2.92&0.23&0.56&0.00058\\
9&16:26:03.6&-24:34:24.0&12.1&10.2&118&0.01&3.49&0.15&0.19&0.00073\\
10&16:26:07.6&-24:19:29.4&87.7&18.3&29&0.04&3.88&0.18&1.24&0.01231\\

\hline
 \end{tabular}
\end{scriptsize}
\label{Table:clumps}
\end{table*}

To estimate the completeness limit, experience with $\it{CPROPS}$ suggests that, for the masking parameters used, the catalogue is complete for clumplets larger than 2 beam areas at the 3$\sigma$ level in the C$^{18}$O data.  The de-blending limit is usually a good deal higher for crowded fields.  However, using the present data set, this gives, for 0.11 K rms on the $T_{\rm {MB}}$ scale, a completeness level given by:

\begin{equation}
F = n_{bm}~n_{\sigma}~\Delta S_{rms}~\Delta v~ PS_{jcmt}
  \label{Eqn:completeness}
\end{equation}

where n$_{bm}$ is the number of beams (taken to be 2 in the present case), $n_{\sigma}$ is the signal-to-noise (taken to be 3$\sigma$), $\Delta S_{rms}$ is the rms noise level in a 0.1 km s$^{-1}$ channel, $\Delta{v}$ is the channel width used in the ${CPROPS}$ analysis (0.1 km s$^{-1}$), and PS$_{jcmt}$ is the point source sensitivity of the JCMT (20.4 Jy K$^{-1}$ beam$^{-1}$). In practice $n_{bm}$ and~$n_{\sigma}$ are both based on the ${CPROPS}$ source identification parameters that are used, and for the present case Equations \ref{Eqn:m_lum} and \ref{Eqn:completeness} then lead to a clump mass completeness of 0.15--0.2~\Msun.

The positions of the molecular C$^{18}$O clumplets reported in Table \ref{Table:clumps} were compared with the seventy-two 850~$\micron$~submm clumps reported by Simpson \al (2008). Of the 105 molecular clumplets, eighteen cross matched with the locations of submm clumps to within 15$^{\prime\prime}$, with the majority of these agreeing to better than 10$^{\prime\prime}$. The dust continuum masses and the gas virial mass estimates in these cases agreed to within factors of $\sim$5 of each other. However, the majority of clumplets do not positionally correlate with the submm clumps, and we have to surmise that the clumplets predominantly trace a different population of material from those exhibiting detectable 850~$\micron$~emission in the Simpson \al (2008) survey.

\begin{figure}
  \centering
  \includegraphics[width=0.99\linewidth]{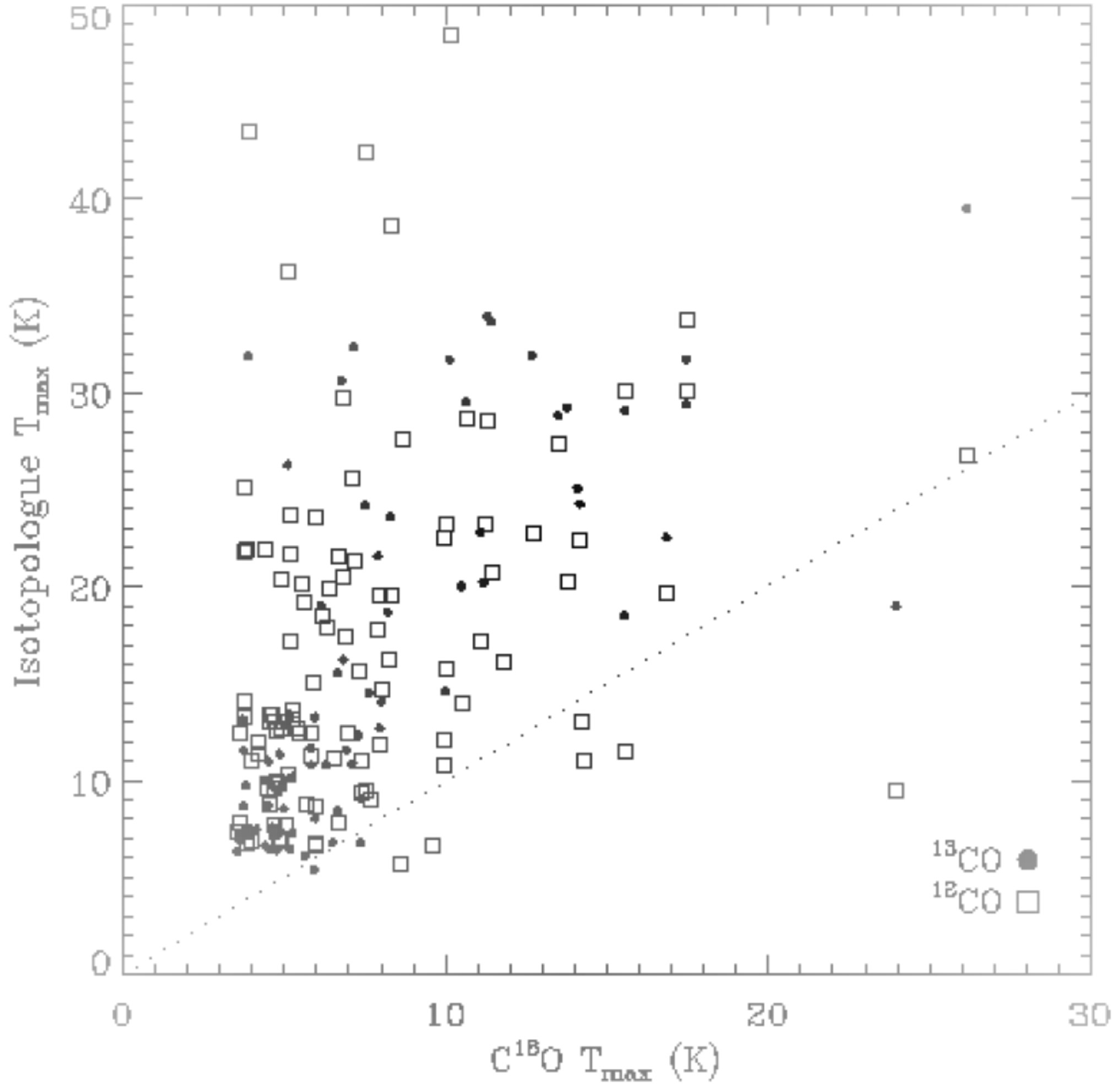}
  \caption{The clumplet excitation temperatures estimated from the CO and $^{13}$CO lines.}
  \label{Fig:coreproperties}
\end{figure}

The thermal properties of the various clumplets detected using ${\it CPROPS}$ are shown in Figure \ref{Fig:coreproperties}, which compares the clump excitation temperatures estimated from the CO, $^{13}$CO and C$^{18}$O lines, after correcting the observed antenna temperatures to main beam brightness temperature.

Figure \ref{Fig:coreproperties} shows that the inferred CO excitation temperatures are generally higher than those inferred from the C$^{18}$O main beam brightness temperatures. This implies that the C$^{18}$O isotopologue may not be fully thermalised, as is consistent with the spectral profiles shown in Figure \ref{COspectra}, or alternatively that the emission is clumpy on sub-beam scales, leading to a reduction in intensity as a consequence of a filling factor that is less than unity.

The Larson relations (Larson 1981) were used to examine the line width-size relation ($\sigma\propto$ R$^{\alpha}$), which relates the turbulent velocity dispersion of a structure to its size via a power law. Although the Larson relationships are known to trace the large-scale turbulent properties of molecular clouds, they may break down on smaller scales due to the increasing dominance of gravity and the transition to coherent (subsonic) linewidths (Goodman \al 1998).

The velocity dispersion obtained from the data is plotted as a function of clumplet radius in Figure \ref{Fig:analysis1}. This can relate the properties of the very small clumplets characterised in this study to those of larger clouds elsewhere in the Galaxy. Although this relationship has been extensively characterised with observations of other molecular clouds, these generally sample clumps with higher radii and dispersion values, and so the current data extend previous work to much smaller size scales. Although the present data show considerable scatter, the smaller clouds appear consistent with that expected for a thermal component $\le$15 K. Despite the fact that the ${\it CPROPS}$ analysis may still suffer from systematic biases remaining in the technique, Figure \ref{Fig:analysis1} suggests that the smaller clumplets may be cooler (as they fall below the predicted 15 K thermal linewidth curve), although about half of the larger clumplets may be more consistent with temperatures in the range 20-30 K. Such a temperature spread is consistent with results in Orion A presented for an ensemble of clumps by Buckle \al (2012). To determine the velocity dispersion, we adopt the following relationship:

\begin{figure}
  \centering
  \includegraphics[width=1.35\linewidth]{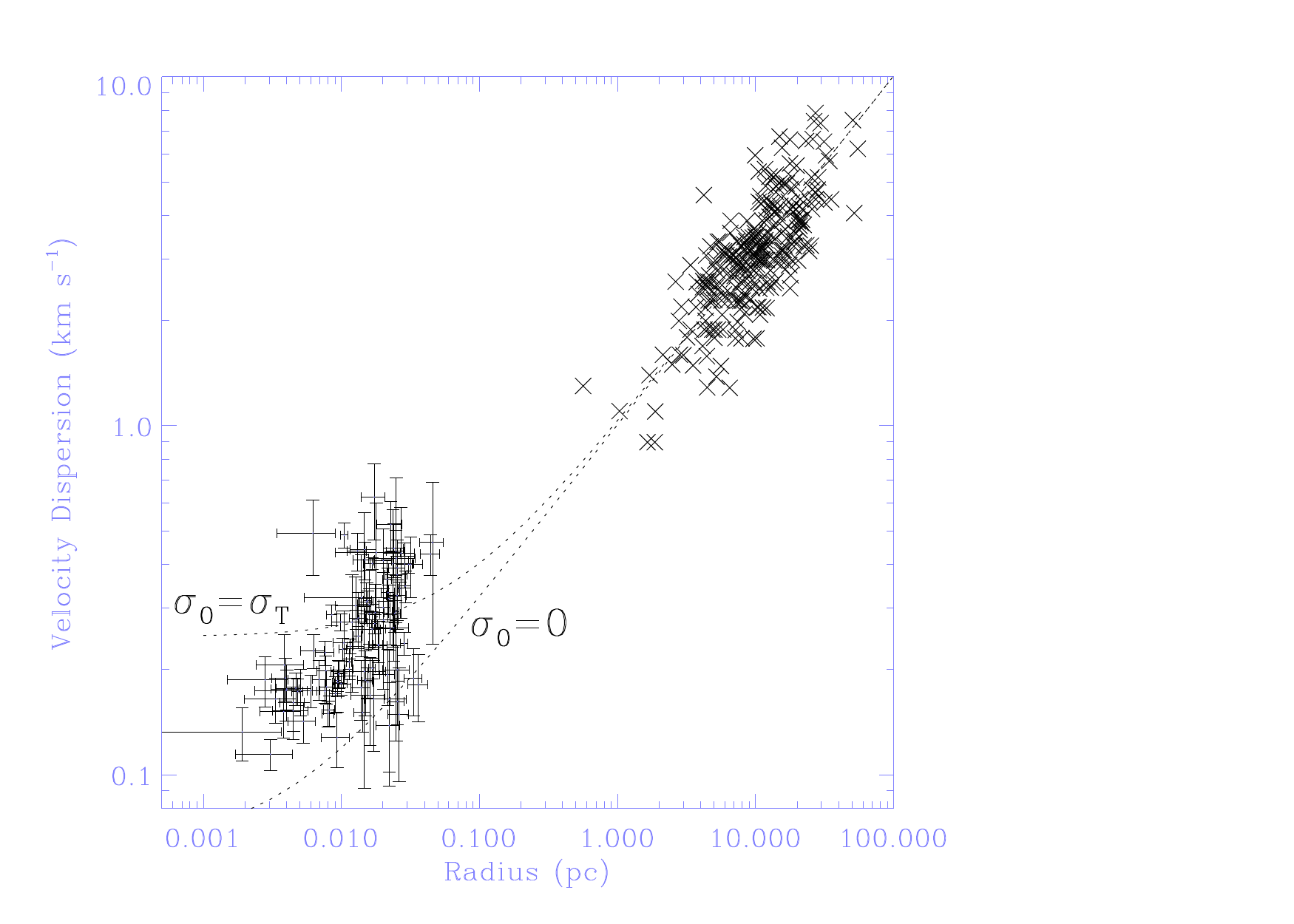}
  \caption{Velocity dispersion-size relationship of the molecular clumplets identified using $\it{CPROPS}$, shown as error bars on the left of the plot. The clumplet radii were estimated as described by Rosolowsky $\&$ Leroy (2006), and have been deconvolved from the telescope beam. The crosses on the right-hand side of the plot are values taken from the Solomon \al (1987) survey of 273 clouds in the Galactic Plane, which showed that these clumps followed a Larson relationship. The two dotted curves show the expected values for the cases where the thermal linewidth, ${\sigma_o}$ is set equal to zero, and for an example curve for an kinetic temperature of 15 K, chosen to be representative of cloud material. At larger values of radii and dispersion, these two curves approach each other, joining to the Solomon \al clouds, and showing (despite some scatter that most plausibly may be a consequence of temperature variations in the clumps) that the Oph clumplets follow extrapolation of the Larson relationship to clump radii below 0.01 pc, when thermal linewidths start to become significant compared to the velocity dispersion. Although the uncertainties are relatively large, many of the Oph CO clumplets have dispersions close to the expected thermal linewidths for temperatures $\sim$ 15 K. $\sigma_{nt}^2$ is therefore always less than 0.1 \kmsec, which is negligible compared to the ${\it average}$ velocity dispersion of C$^{18}$O in Ophiuchus, which is $\sim$0.8~\kmsec. This is similar to that inferred for the W51 clump sample (Parsons \al 2012), and slightly lower than the mean value inferred for clumps toward Orion A (Buckle \al 2012).}
  \label{Fig:analysis1}
\end{figure}

\begin{equation}
\sigma_{tot} = \sqrt{\sigma_{nt}^2+\sigma_{o}^2(\mu)}
\end{equation}

where $\sigma_{nt}^2$ and $\sigma_{o}^2(\mu)$ are the non-thermal and thermal linewidths respectively, with the thermal component being given by:

\begin{equation}
\sigma_{o}^2(\mu)=\sqrt{\frac{kT}{\mu m_H}}
\end{equation}

with $m_H$ the mass of a hydrogen atom, and $\mu$ the mean molecular weight of C$^{18}$O $(\mu = 30)$. 

The data are over-plotted with curves showing the expected variation with $\sigma_{o}$ for values of 0 and 15 K. The data are broadly consistent with this range, given that there will naturally be a spread in the gas temperatures from clump to clump. In general the size-line width trend observed is consistent with the standard GMC based size-line width relationship, with the clumplets having higher line widths as showing a similar trend to higher pressure regions of the galaxy.

\begin{figure}
  \centering
  \includegraphics[width=0.99\linewidth]{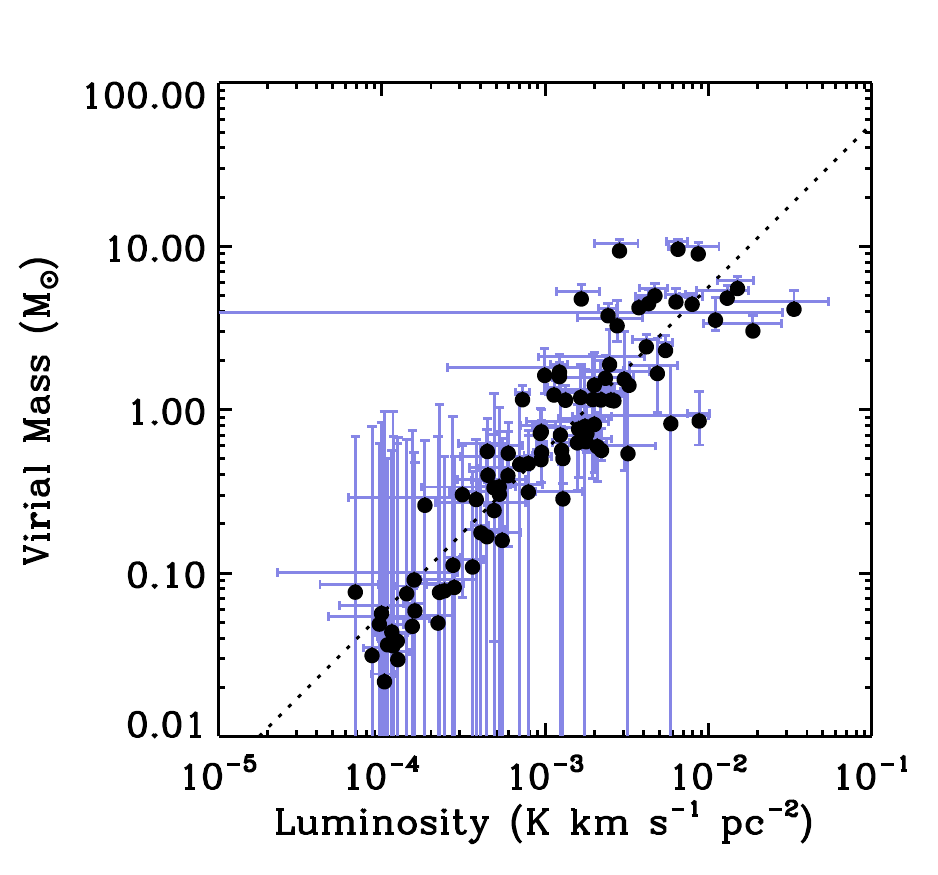}
  \caption{Virial masses (in M$_{\rm sun}$) of the molecular clumplets identified by $\it{CPROPS}$ plotted against the CO luminosity (in K \kmsec~pc$^{-2}$). The dotted line shows shows that expected if $M_{vir}$ = $M_{lum}$ using a CO-to-H$_2$ conversion factor of 4.95~$\times$~10$^{22}$~cm$^2$ ( K km s$^{-1}$)$^{-1}$, which is established as the line that goes through those points. The clumplets detected here follow a continuous relationship, that with other data in the literature extends over almost 9 orders of magnitude in our own Galaxy, and extends out to local group galaxies with appropriate scaling of X$_{CO}$ for metallicity (Bolatto \al 2008, Hughes \al 2010).}
  \label{Fig:analysis2}
\end{figure}

The correlation between the dynamical (virial) and luminous masses of the molecular clumplets is examined in Figure \ref{Fig:analysis2}. Although many of the points whose virial masses are below $\sim$ 1~\Msun~have relatively large uncertainties (see Section \ref{clump_analysis} for discussion of s/n limitations), these points have been left on Figure \ref{Fig:analysis2} with their mean values indicated by filled circles as they do appear to follow the trend. The luminosity for the x-axis and the sizes and linewidths that contribute to the virial mass are all deduced by moments of the intensity distribution.  These moments have been extrapolated down to their expected values in the absence of noise (as described by Rosolowsky $\&$ Leroy 2006).  To assess the significance of an error bar, the intensity distribution (i.e. a set of (x, y, v, T$_a^*$) data) was sampled with replication to draw the same number of points from the data as in the original clump, thus enabling extrapolation of the points down to the zero level.  This is a 'bootstrap' approach, which has been theoretically justified as an empirical estimate of uncertainties given a set of measurements (Efron 1979). This procedure was repeated 1000 times for each clump and the error distribution measures the width of the resulting distribution of samples.  This method has been shown in practice to do a good job in assessing the uncertainties from noise in the data and from the extrapolation process.  It is however less good at assessing the systematic errors engendered by the cataloguing method itself, which is a more difficult problem to solve rigorously. On Figure \ref{Fig:analysis2}, those points falling above the dotted line are most likely gravitationally unbound or pressure confined, whereas those falling below the dotted line may be self-gravitating. Despite the rather large uncertainties, it does appear that at least some of the identified clumplets may be able to collapse to form stars.

Figure \ref{Fig:analysis2} shows an approximately linear correlation between a dynamical estimate of the mass (i.e. the virial mass estimated assuming uniform density) compared to the luminosity of the line (Bolatto \al 2008). Clumplets undergoing gravitational collapse would lie on the right-hand side of the dotted line, whereas pressure-confined or gravitationally unbound clumplets would lie to the left. The majority of the data points cluster closely to the line between these two constraints. Within the noise constraints, there are a small number of clumplets which may be in free-fall, as Simpson \al 2011 have suggested), or alternatively may be pressure confined. Extrapolation of the dotted line to higher M$_{vir}$ and luminosity ranges shows that this linear trend holds over 11 dex in luminosity (see for example Rebolledo \al 2012).

\section{Summary}
\label{Section:Summary}
We have analysed JCMT Gould Belt survey CO, $^{13}$CO and C$^{18}$O~3--2 maps of the main star-forming cloud in Ophiuchus ~LDN1688. Our main conclusions are:
\begin{enumerate}
\item In Oph~A, the typical integrated intensity ratios $^{13}$CO/C$^{18}$O range from $\sim$2--7, corresponding to $\tau (\mathrm{^{13}CO})$ of 2--16 and $\tau(\mathrm{C^{18}O})$ of $\sim$0.1--2.0.  $^{13}$CO is usually found to be optically thick around the main clumps.  C$^{18}$O is generally optically thin across 96 per cent of the whole area mapped, although does reach opacities up to 2 in several of the clumps.  In particular the more abundant $^{13}$CO emission from the PDR region is self-absorbed with a $^{13}$CO/C$^{18}$O integrated intensity ratio as low as 1.2.
\item From the isotopomeric lines, the gravitational binding energy is estimated to be 4.5 $\times$ 10$^{39}$ J (3497~\Msun~km$^2$ s$^{-2}$).
\item The velocity structure in the gas around protostars originally detected by the \emph{Spitzer} c2d survey data were examined. Outflows were detected towards 28 out of 30 sources (8 firm associations, with 20 more marginal).  Several new sources identified in the c2d survey have evidence for red and/or blueshifted outflow lobes, and it is less likely that flat spectrum objects have firm outflow detections than is the case for Class 0/I protostars.
\item After inclination and optical depth are corrected for, the total outflow kinetic energy is 1.3 $\times$ 10$^{38}$~J (67~$M_\odot$~km$^2$~s$^{-2}$) corresponding to 21~per cent of the turbulent kinetic energy.  This suggests that outflows are significant, but ${\it not}$ the dominant source of turbulence in Ophiuchus.
\item  The turbulent energy is estimated to be 6.3 $\times$ 10$^{38}$ J (3497~\Msun~km$^2$ s$^{-2}$), which is roughly a factor of 7 times smaller than the binding energy, suggesting both that the cloud as a whole is gravitationally bound, and also that the outflows are not the dominant source of turbulent injection into the Oph cloud.
\item That the turbulent kinetic energy is too small to support the cloud, and the outflow kinetic energy is insufficient to be the dominant driver of turbulence (since the outflows are also unable to support the cloud), leads to the inevitable conclusion that on the large scale, the Ophiuchus cloud is in a state of collapse.
\item 105 clumplets were detected within the cloud using the $\it{CPROPS}$ source extraction algorithm, whose radii ranged from 0.01--0.05 pc, and which had virial masses between 0.1 and 12~\Msun, and luminosities between 0.001 and 0.1 K~km~s$^{-1}$~pc$^{-2}$. Eighteen of these lay within one beamwidth of a submm 850~$\micron$~core.
\item The excitation temperatures of the clumplets ranged from 10--50 K, with the majority of the warmest clumplets being associated with the Oph A region, and from close to the PDR..
\item From the velocity dispersion--radius plot, the expected variation is well fitted for clumplets having a typical temperature of $\sim$10--25 K, with the $\sigma_{o}$ set equal to zero, or to its thermal width. In general the size-line width trend observed is consistent with the standard GMC based size-line width relationship, with the clumplets having lower line widths than that of larger molecular clouds.
\item Comparison between the virial mass, as a dynamical estimate of the masses of individual clumplets, and the luminosity of the cloud clumplet ensemble shows a good correlation consistent with a C$^{18}$O-to-H$_2$ conversion factor of 4.95 $\times$ 10$^{22}$ cm$^2$ / K \kmsec, although a small number of clumplets may either be in free-fall, or be pressure confined.

\end{enumerate}

\section{Acknowledgements}

We thank Alain Abergel for sharing his archival ISOCAM data at an early stage of writing this paper. The JCMT is operated by the Joint Astronomy Centre, Hawaii, on behalf of the UK STFC, the Netherlands NWO and the Canadian NRC. We gratefully acknowledge the support of the JCMT staff and operators.

\onecolumn

\newcounter{tablemajor}
\renewcommand\thetable{\arabic{tablemajor}}
\newcommand*\settablecounter[1]{%
        \setcounter{tablemajor}{#1}%
}

\settablecounter{5}{}
\begin{center}
\begin{longtable}{|l|l|l|l|l|l|l|l|l|l|l}
\caption[The clumplet catalogue. The clumplets listed in the catalogue are: (1) a short form running number, (2,3) Right Ascension and Declination (J2000) of the clump centre, (4,5,6) the major and minor axis of the clump deconvolved from the beam in arc seconds, along with the position angle (measured east of north), (7) the deconvolved radius of the clump as defined by Rosolowsky \al (2006), (8,9) centre v$_{LSR}$ and dispersion of the clump as determined from the C$^{18}$O data, (10) The virial mass, and (11) the Luminous mass estimated from the CO observations.]{The clumplet catalogue. The clumplets listed in the catalogue are: (1) a short form running number, (2,3) Right Ascension and Declination (J2000) of the clump centre, (4,5,6) the major and minor axis of the clump deconvolved from the beam in arc seconds, along with the position angle (measured east of north), (7) the deconvolved radius of the clump as defined by Rosolowsky \al (2006), (8,9) centre v$_{LSR}$ and dispersion of the clump as determined from the C$^{18}$O data, (10) The virial mass, and (11) the Luminous mass estimated from the CO observations.} 
\label{grid_mlmmh} \\

\multicolumn{1}{r}{No} & \multicolumn{1}{c}{RA (J2000)} & \multicolumn{1}{c}{Dec (J2000)} & \multicolumn{1}{c}{Maj} & \multicolumn{1}{c}{Min} & \multicolumn{1}{c}{Pos angle} & \multicolumn{1}{c}{Radius} & \multicolumn{1}{c}{Velocity} & \multicolumn{1}{c}{Dispersion} & \multicolumn{1}{c}{Virial mass} & \multicolumn{1}{c}{Luminosity} \\ 
 \multicolumn{1}{l}{} & \multicolumn{1}{c}{h:m:s.s}   & \multicolumn{1}{c}{d:m:s.s} &  \multicolumn{1}{c}{${\prime\prime}$} & \multicolumn{1}{c}{${\prime\prime}$} &  & \multicolumn{1}{c}{pc} & \multicolumn{1}{c}{km s$^{-1}$} & \multicolumn{1}{c}{km s$^{-1}$}  & \multicolumn{1}{c}{\Msun}  & \multicolumn{1}{c}{K km s$^{-1}$ pc$^{-2}$} \\
\multicolumn{1}{l}{(1)} & \multicolumn{1}{c}{(2)} & \multicolumn{1}{c}{(3)} & \multicolumn{1}{c}{(4)} &  \multicolumn{1}{c}{(5)} & \multicolumn{1}{c}{(6)} & \multicolumn{1}{c}{(7)} & \multicolumn{1}{c}{(8)} & \multicolumn{1}{c}{(9)} & \multicolumn{1}{c}{(10)} & \multicolumn{1}{c}{(11)}  \\
\hline 
\endfirsthead
\hline
\endhead


\hline \hline
\endlastfoot

1&16:25:54.0&-24:29:46.7&22.9&16.1&51&0.02&2.95&0.20&0.71&0.00141\\
2&16:25:55.4&-24:17:30.2&7.9&7.7&-38&0.00&3.54&0.11&0.04&0.00015\\
3&16:25:56.9&-24:23:25.3&12.8&11.2&-41&0.01&3.45&0.17&0.27&0.00034\\
4&16:26:00.1&-24:31:24.7&42.1&13.2&2&0.02&3.41&0.28&1.62&0.00147\\
5&16:26:00.4&-24:33:53.0&8.4&8.0&63&0.00&3.69&0.17&0.13&0.00022\\
6&16:26:01.3&-24:32:29.1&26.6&14.2&132&0.02&3.35&0.26&1.18&0.00146\\
7&16:26:02.1&-24:18:45.2&11.0&27.2&65&0.01&3.75&0.32&1.51&0.00781\\
8&16:26:02.5&-24:25:31.2&13.6&12.4&118&0.01&2.92&0.23&0.56&0.00058\\
9&16:26:03.6&-24:34:24.0&12.1&10.2&118&0.01&3.49&0.15&0.19&0.00073\\
10&16:26:07.6&-24:19:29.4&87.7&18.3&29&0.04&3.88&0.18&1.24&0.01231\\
11&16:26:09.3&-24:17:54.0&43.8&16.2&92&0.02&3.17&0.40&3.99&0.00440\\
12&16:26:09.9&-24:19:40.6&31.4&19.3&54&0.02&2.72&0.32&2.34&0.00532\\
13&16:26:10.1&-24:23:21.0&10.2&9.2&72&0.01&3.41&0.49&1.57&0.00073\\
14&16:26:10.9&-24:24:43.7&35.1&21.5&67&0.02&2.43&0.44&4.93&0.00334\\
15&16:26:12.2&-24:15:24.5&9.3&8.7&0&0.01&2.87&0.14&0.11&0.00027\\
16&16:26:12.3&-24:17:21.6&18.9&32.6&0&0.02&2.22&0.14&0.44&0.00457\\
17&16:26:12.4&-24:32:11.5&8.4&7.6&86&0.00&3.45&0.16&0.09&0.00028\\
18&16:26:13.3&-24:22:05.3&25.3&12.4&123&0.01&4.27&0.41&2.62&0.00094\\
19&16:26:13.9&-24:21:37.9&143.0&18.0&-26&0.05&3.20&0.46&10.19&0.00626\\
20&16:26:14.3&-24:25:10.7&7.9&8.5&73&0.00&4.46&0.21&0.17&0.00010\\
21&16:26:14.7&-24:25:34.2&57.2&39.6&61&0.04&3.21&0.43&8.46&0.00456\\
22&16:26:16.0&-24:14:50.8&8.2&8.0&33&0.00&2.68&0.15&0.09&0.00013\\
23&16:26:16.4&-24:23:57.6&18.8&11.0&102&0.01&4.06&0.20&0.48&0.00061\\
24&16:26:16.7&-24:25:31.3&25.8&11.2&3&0.01&2.49&0.18&0.46&0.00145\\ 
25&16:26:18.3&-24:14:24.4&11.0&7.5&17&0.00&3.04&0.15&0.11&0.00013\\
26&16:26:19.1&-24:22:39.5&48.9&16.5&-22&0.03&2.63&0.37&3.68&0.00564\\
27&16:26:20.8&-24:33:51.7&17.0&11.8&132&0.01&2.62&0.21&0.52&0.00093\\
28&16:26:21.7&-24:26:45.7&23.9&13.8&0&0.02&2.66&0.32&1.61&0.00101\\
29&16:26:21.9&-24:16:24.2&9.1&7.7&119&0.00&3.60&0.18&0.13&0.00018\\
30&16:26:22.4&-24:19:53.7&50.9&14.7&77&0.02&3.37&0.45&4.96&0.00526\\
31&16:26:23.5&-24:20:29.3&16.9&12.7&50&0.01&2.80&0.27&0.93&0.00143\\
32&16:26:25.9&-24:32:54.9&12.0&9.7&26&0.01&3.21&0.17&0.24&0.00046\\
33&16:26:26.1&-24:24:51.0&10.5&8.8&114&0.01&1.90&0.17&0.19&0.00025\\
34&16:26:27.6&-24:23:52.4&22.5&18.3&77&0.02&3.65&0.43&3.49&0.01874\\
35&16:26:27.7&-24:14:58.9&43.0&17.8&27&0.02&2.40&0.29&2.22&0.00312\\
36&16:26:27.8&-24:26:44.5&16.4&12.4&43&0.01&3.93&0.23&0.66&0.00072\\
37&16:26:28.4&-24:22:56.1&24.6&12.5&103&0.01&2.88&0.44&2.98&0.00975\\
38&16:26:31.2&-24:18:53.1&59.4&15.7&-19&0.03&3.03&0.43&5.22&0.00225\\
39&16:26:31.6&-24:18:42.0&25.1&16.1&-8&0.02&3.85&0.29&1.55&0.00162\\
40&16:26:32.2&-24:16:02.2&24.2&16.2&-6&0.02&3.17&0.62&7.02&0.00152\\
41&16:26:32.5&-24:26:17.1&62.4&19.7&54&0.03&2.76&0.40&5.30&0.01337\\
42&16:26:33.2&-24:24:16.3&17.4&9.3&30&0.01&2.81&0.13&0.16&0.00070\\
43&16:26:34.6&-24:28:08.8&61.6&13.8&33&0.03&2.66&0.39&4.04&0.00568\\
44&16:26:36.2&-24:17:58.5&32.5&22.3&-7&0.02&3.54&0.29&2.13&0.00396\\
45&16:26:36.9&-24:20:18.3&7.5&7.2&107&0.00&2.79&0.13&0.04&0.00029\\
46&16:26:39.7&-24:31:41.8&15.3&10.5&-8&0.01&2.93&0.19&0.38&0.00063\\
47&16:26:41.8&-24:24:02.7&12.3&8.8&-22&0.01&3.21&0.19&0.25&0.00041\\
48&16:26:41.8&-24:17:44.1&27.7&20.7&38&0.02&2.50&0.36&2.94&0.00422\\
49&16:26:44.3&-24:34:40.6&11.6&8.5&-24&0.01&4.10&0.23&0.33&0.00020\\
50&16:26:45.4&-24:19:12.9&16.4&9.8&6&0.01&3.78&0.18&0.33&0.00036\\
51&16:26:46.1&-24:29:13.9&32.2&14.3&54&0.02&2.91&0.23&1.07&0.00235\\
52&16:26:48.3&-24:20:13.6&39.7&10.2&130&0.02&2.89&0.29&1.39&0.00067\\
53&16:26:49.1&-24:32:02.0&9.0&8.7&84&0.01&2.95&0.17&0.16&0.00017\\
54&16:26:50.4&-24:23:27.3&20.5&12.4&-20&0.01&2.76&0.25&0.86&0.00136\\
55&16:26:51.9&-24:35:16.5&7.5&6.8&100&0.01&3.84&0.14&0.35&0.00003\\
56&16:26:52.6&-24:20:02.8&25.9&10.6&56&0.01&3.11&0.28&1.09&0.00103\\
57&16:26:52.6&-24:33:28.0&12.1&9.8&68&0.01&4.44&0.20&0.32&0.00021\\
58&16:26:56.6&-24:34:40.6&16.1&11.1&6&0.01&3.08&0.49&2.58&0.00047\\
59&16:26:58.2&-24:37:46.6&33.8&23.0&53&0.03&4.06&0.33&2.80&0.00168\\
60&16:26:59.6&-24:22:27.0&26.6&12.8&-18&0.02&2.72&0.18&0.55&0.00088\\
61&16:26:59.7&-24:29:16.9&13.8&9.5&56&0.01&2.80&0.15&0.20&0.00034\\
62&16:26:59.7&-24:35:12.9&10.0&7.5&45&0.00&3.51&0.19&0.15&0.00010\\
63&16:27:00.7&-24:22:40.4&38.1&20.8&-21&0.03&3.26&0.16&0.70&0.00084\\
64&16:27:02.7&-24:38:39.8&34.9&17.8&37&0.02&4.45&0.16&0.63&0.00160\\
65&16:27:05.5&-24:29:35.1&12.9&9.5&116&0.01&2.62&0.18&0.26&0.00023\\
66&16:27:06.2&-24:38:17.6&21.3&16.3&57&0.02&3.61&0.17&0.48&0.00032\\
67&16:27:07.2&-24:32:49.1&13.6&12.7&96&0.01&2.59&0.24&0.62&0.00039\\
68&16:27:07.4&-24:37:02.2&54.7&12.2&59&0.02&5.18&0.33&2.51&0.00186\\
69&16:27:07.8&-24:28:20.7&56.2&12.7&73&0.02&3.82&0.52&6.45&0.00250\\
70&16:27:08.0&-24:39:39.4&29.9&26.3&18&0.03&4.09&0.20&1.06&0.00211\\
71&16:27:09.9&-24:27:19.3&9.5&8.0&43&0.00&2.64&0.18&0.15&0.00011\\
72&16:27:10.2&-24:39:22.8&37.2&10.1&26&0.02&3.22&0.31&1.55&0.00081\\
73&16:27:10.5&-24:30:28.2&53.4&19.0&81&0.03&3.12&0.40&4.81&0.00205\\
74&16:27:10.7&-24:26:59.0&81.4&11.9&128&0.03&4.19&0.15&0.61&0.00215\\
75&16:27:10.9&-24:30:19.6&16.4&17.0&81&0.01&3.74&0.15&0.34&0.00063\\
76&16:27:11.4&-24:32:07.9&22.3&19.6&-17&0.02&3.42&0.24&1.07&0.00099\\
77&16:27:11.5&-24:38:07.9&33.5&19.0&38&0.02&4.62&0.26&1.61&0.00206\\
78&16:27:12.5&-24:25:20.7&50.0&20.0&121&0.03&3.37&0.24&1.69&0.00180\\
79&16:27:17.4&-24:40:58.2&19.7&15.0&0&0.01&3.16&0.20&0.60&0.00130\\
80&16:27:17.6&-24:42:03.6&51.7&20.0&53&0.03&3.92&0.35&3.65&0.00197\\
81&16:27:19.1&-24:31:42.2&25.7&12.4&95&0.01&4.20&0.19&0.54&0.00052\\
82&16:27:19.4&-24:29:09.6&49.5&22.8&103&0.03&4.40&0.40&5.12&0.00373\\
83&16:27:20.3&-24:39:51.3&19.9&12.1&39&0.01&4.55&0.30&1.23&0.00112\\
84&16:27:21.5&-24:26:32.0&24.5&15.2&-18&0.02&4.35&0.40&2.82&0.00117\\
85&16:27:22.2&-24:39:46.5&29.7&25.5&70&0.02&3.73&0.42&4.52&0.00182\\
86&16:27:23.8&-24:26:52.2&16.0&10.0&22&0.01&3.05&0.19&0.38&0.00025\\
87&16:27:24.1&-24:23:59.1&47.4&18.4&30&0.03&2.98&0.34&3.22&0.00215\\
88&16:27:26.4&-24:33:17.5&8.9&8.2&38&0.00&3.60&0.16&0.12&0.00012\\
89&16:27:27.6&-24:39:24.2&15.3&15.8&63&0.01&3.23&0.44&2.61&0.00091\\
90&16:27:28.0&-24:40:49.3&56.4&23.5&58&0.03&4.15&0.19&1.23&0.00151\\
91&16:27:29.9&-24:28:06.4&21.8&8.8&117&0.01&2.79&0.27&0.77&0.00028\\
92&16:27:30.4&-24:27:39.3&13.2&10.2&134&0.01&4.65&0.29&0.73&0.00044\\
93&16:27:30.9&-24:29:35.1&14.8&7.3&52&0.00&3.76&0.17&0.15&0.00011\\
94&16:27:37.0&-24:41:58.0&22.1&17.4&-16&0.02&3.97&0.16&0.48&0.00055\\
95&16:27:37.5&-24:27:00.1&37.3&16.1&44&0.02&3.27&0.19&0.84&0.00109\\
96&16:27:38.9&-24:27:07.0&34.1&15.3&62&0.02&3.84&0.30&1.89&0.00140\\
97&16:27:38.9&-24:35:25.9&14.8&8.7&46&0.01&2.76&0.22&0.40&0.00024\\
98&16:27:40.1&-24:24:37.4&26.3&19.4&104&0.02&3.20&0.41&3.53&0.00042\\
~99&16:27:40.6&-24:41:57.6&31.1&22.6&75&0.02&3.61&0.26&1.74&0.00213\\
100&16:27:41.1&-24:31:60.0&7.3&8.6&78&0.00&4.27&0.19&0.10&0.00011\\
101&16:27:42.4&-24:24:34.7&28.1&12.7&34&0.02&3.73&0.31&1.57&0.00054\\
102&16:27:44.0&-24:37:18.6&20.5&11.9&118&0.01&3.18&0.28&1.03&0.00036\\
103&16:27:52.4&-24:31:40.0&16.2&9.8&49&0.01&4.07&0.19&0.34&0.00036\\
104&16:27:53.8&-24:34:53.2&14.1&11.6&112&0.01&4.16&0.18&0.34&0.00034\\
105&16:28:05.3&-24:36:09.4&25.0&16.0&-12&0.02&3.64&0.27&1.33&0.00076\\
\end{longtable}
\end{center}

\end{document}